\documentclass[numberedappendix]{emulateapj}
\usepackage{natbib}
\usepackage{graphicx}
\usepackage{dcolumn}
\usepackage{bm}
\usepackage{amsmath}
\usepackage{amstext}
\usepackage{pstricks}
\usepackage{color}
\renewcommand{\baselinestretch}{1.18}

\newcommand\bp{\begin{figure}}
\newcommand\ep{\end{figure}}
\newcommand\bpm{\begin{figure*}}
\newcommand\epm{\end{figure*}}
\newcommand\reffig[1]{Figure \ref{fig:#1}}

\newcommand\refsec[1]{\S \ref{sec:#1}}
\newcommand\Refsec[1]{Section \ref{sec:#1}}
\newcommand\reftbl[1]{Table \ref{tbl:#1}}

\newcommand{\sigmav}{\langle\sigma v\rangle}

\newcommand{\kevflux}{\rm \, keV\,cm^{-2} s^{-1} sr^{-1}}

\newcommand{\be}{\begin{equation}}
\newcommand{\ee}{\end{equation}}
\newcommand{\bea}{\begin{eqnarray}}
\newcommand{\eea}{\end{eqnarray}}

\newcommand{\degree}{^\circ}

\newcommand{\Fermi}{\emph{Fermi}}

\newcommand{\fb}{\emph{Fermi}\,bubbles\,}

\renewcommand\refname{References and Notes}

\begin{document}

\title{Strong Evidence for Gamma-ray Line Emission from the Inner Galaxy}

\author{Meng Su\altaffilmark{1,3}, Douglas P. Finkbeiner\altaffilmark{1,2}}

\altaffiltext{1}{ 
  Institute for Theory and Computation,
  Harvard-Smithsonian Center for Astrophysics, 
  60 Garden Street, MS-51, Cambridge, MA 02138 USA } 

\altaffiltext{2}{ 
  Physics Department, 
  Harvard University, 
  Cambridge, MA 02138 USA }
\altaffiltext{3}{mengsu@cfa.harvard.edu}








\begin{abstract}
  Using 3.7 years of \Fermi-LAT data, we examine the diffuse $80-200$ GeV
  emission in the inner Galaxy and find a resolved gamma-ray feature at $\sim
  110-140$ GeV.  We model the spatial distribution of this emission with a
  $\sim3\degree$ FWHM Gaussian, finding a best fit position $1.5\degree$ West of the Galactic
  Center.  Even better fits are obtained for off-center Einasto and power-law
  profiles, which are preferred over the null (no line) hypothesis
  by $6.5\sigma$ ($5.0\sigma$/$5.4\sigma$ after trials factor correction for
  one/two line case) assuming an NFW density profile centered at $(\ell,
  b)=(-1.5\degree,0\degree)$ with a power index $\alpha=1.2$ .  The energy
  spectrum of this structure is consistent with a single spectral line (at
  energy $127.0\pm 2.0$ GeV with $\chi^2=4.48$ for 4 d.o.f.).  A pair of lines
  at $110.8\pm 4.4$ GeV and $128.8\pm 2.7$ GeV provides a marginally better
  fit (with $\chi^2=1.25$ for 2 d.o.f.).  The total luminosity of the
  structure is $(3.2\pm0.6)\times 10^{35}$ erg/s, or $(1.7\pm0.4)\times
  10^{36}$ photons/sec.  The energies in the two-line case are compatible with
  a $127.3 \pm 2.7$ GeV WIMP annihilating through $\gamma \gamma$ and $\gamma
  Z$ (with $\chi^2=1.67$ for 3 d.o.f.).  
  We describe a possible change to the \Fermi\
  scan strategy that would accumulate S/N on spectral lines in
  the Galactic center 4 times as fast as the current survey
  strategy.



\end{abstract}


\section{Introduction}
\label{sec:introduction}

Although various cosmological and astrophysical observations
provide compelling evidence for dark matter (DM), which
constitutes $\sim 80\%$ of the matter in the Universe, we
still know little about its intriguing
nature~\citep[e.g.][]{Bertone:2005,Hooper:2007Review}. Among
a forest of dark matter models, stable Weakly Interacting
Massive Particles (WIMP) have been predicted in many
extensions of the Standard Model of particle
physics~\citep[e.g.][]{Bergstrom:2000}. The WIMP with the
virtues of weak scale masses and couplings is an excellent
dark matter particle candidate which can annihilate into
high energy gamma-rays~\citep[e.g.][]{Bergstrom:1998}. The
inner Galaxy provides one of the most promising regions on
the sky to search for WIMP annihilation produced gamma
rays~\citep{Ackermann:2011,Abramowski:2011,Gondolo:1999}. The
expected relatively higher annihilation rate due to higher
dark matter particle density provides a potential window to
identify any non-gravitational dark matter signatures.

\begin{figure*}[ht]
    \begin{center}
        \includegraphics[width=0.8\textwidth]{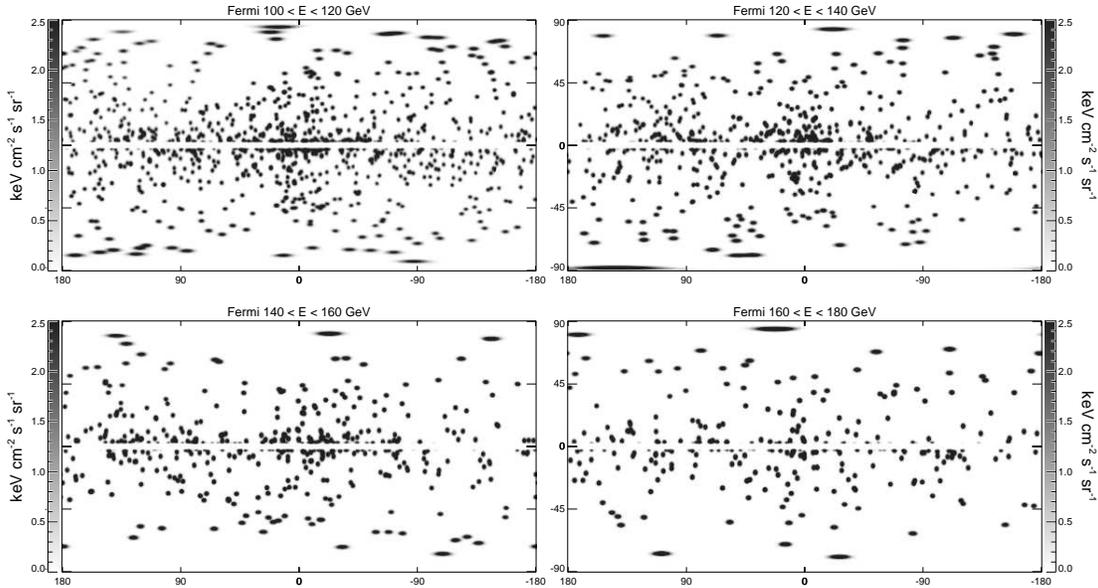}
    \end{center}
\caption{All-sky \Fermi-LAT 3.7 year sky maps in 4 energy bins ranging from 100 to 180 GeV. We
use \texttt{CLEAN} event class and point sources have been
subtracted based on the Second \Fermi-LAT catalog (2FGL).
Large sources, including the inner disk ($-2\degree < b
< 2\degree, -180\degree < \ell < 180\degree$), have been
masked. The maps have been smoothed for display with a
Gaussian kernel of FWHM = 2$\degree$.}
\label{fig:fig1}
\end{figure*}

\begin{figure*}[ht]
    \begin{center}
        \includegraphics[width=0.8\textwidth]{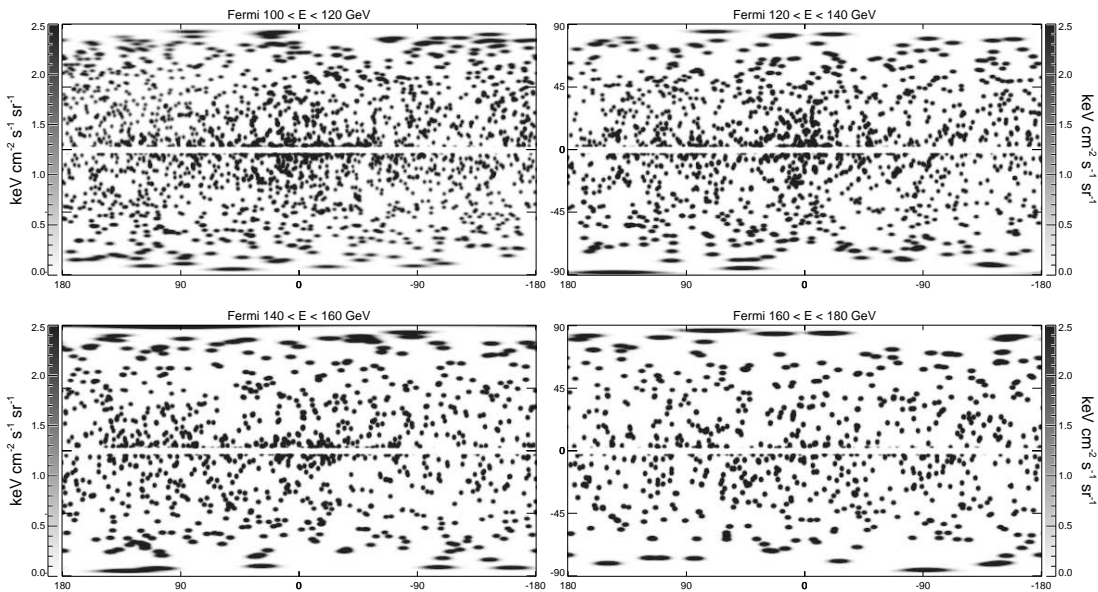}
    \end{center}
\caption{Same as \reffig{fig1}, but with \texttt{SOURCE} class events.  This
  event class contains substantially more background.}
\label{fig:fig2}
\end{figure*}

\begin{figure*}[ht]
  \begin{center}
    \includegraphics[width=0.8\textwidth]{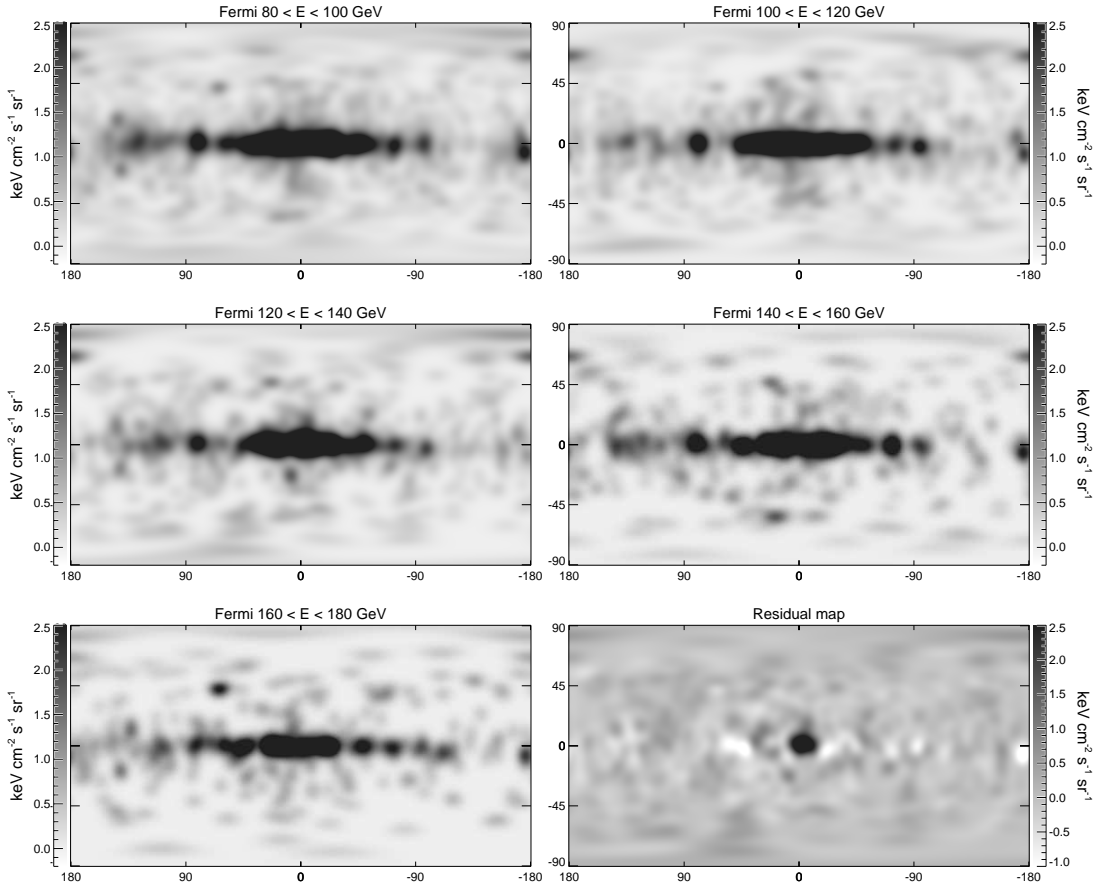}
  \end{center}
  \caption{All-sky \texttt{CLEAN} 3.7 year maps in 5 energy bins, and a
    residual map (\emph{lower right}).  The residual map is the $120-140$ GeV
    map minus a background estimate, taken to be the average of the other 4
    maps where the average is computed in $E^2 dN/dE$ units.  This simple
    background estimate is sufficient to remove the Galactic plane and most of
    the large-scale diffuse structures and even bright point sources.  A cuspy
    structure toward the Galactic center is revealed as the only significant
    structure in the residual gamma-ray map.  All of the maps are smoothed
    with a Gaussian kernel of FWHM = $10\degree$ without source subtraction.  }
  \label{fig:fig3}
\end{figure*}

\begin{figure*}[ht]
    \begin{center}
        \includegraphics[width=0.8\textwidth]{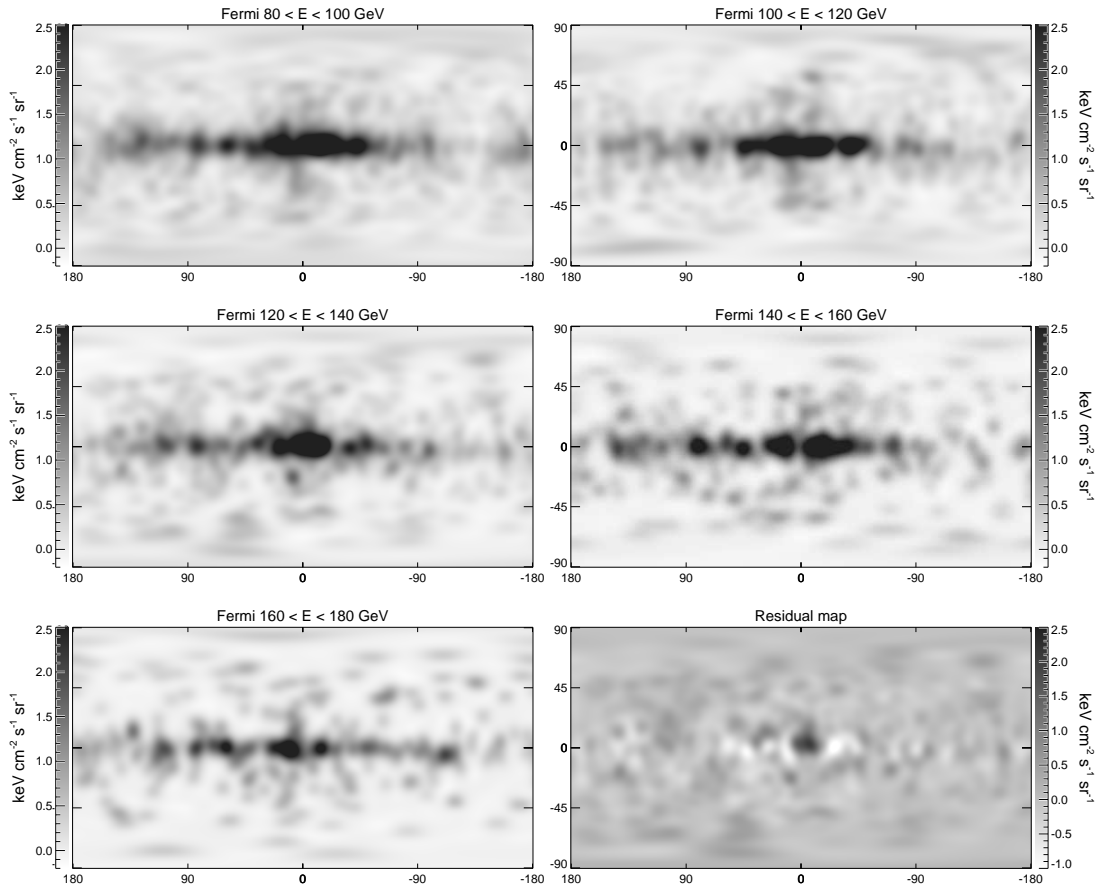}
    \end{center}
\caption{The same as \reffig{fig3} but we have subtracted
point sources before smooth the maps.  }
\label{fig:fig4}
\end{figure*}

\begin{figure*}[ht]
    \begin{center}
        \includegraphics[width=0.47\textwidth]{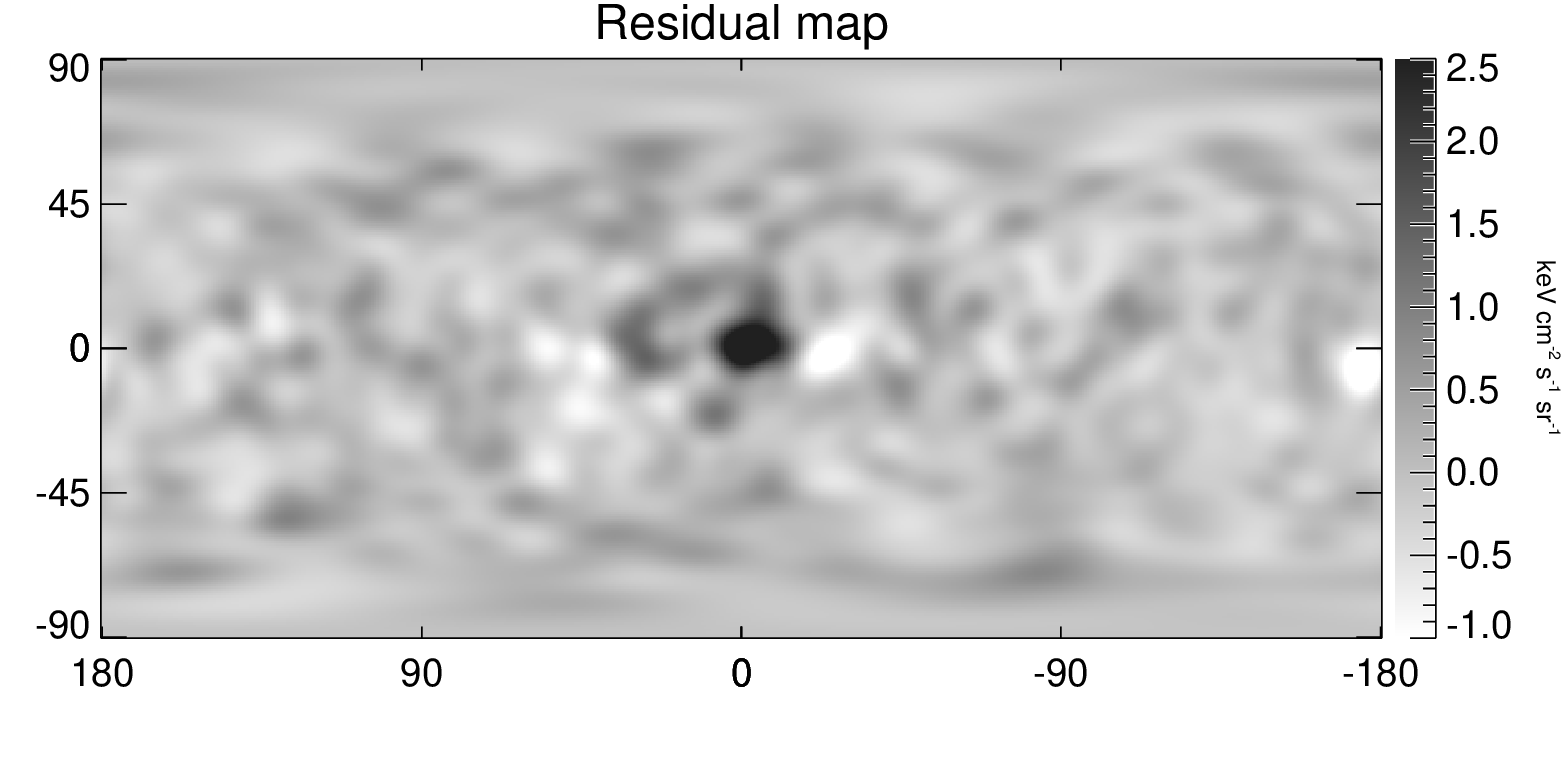}
        \includegraphics[width=0.47\textwidth]{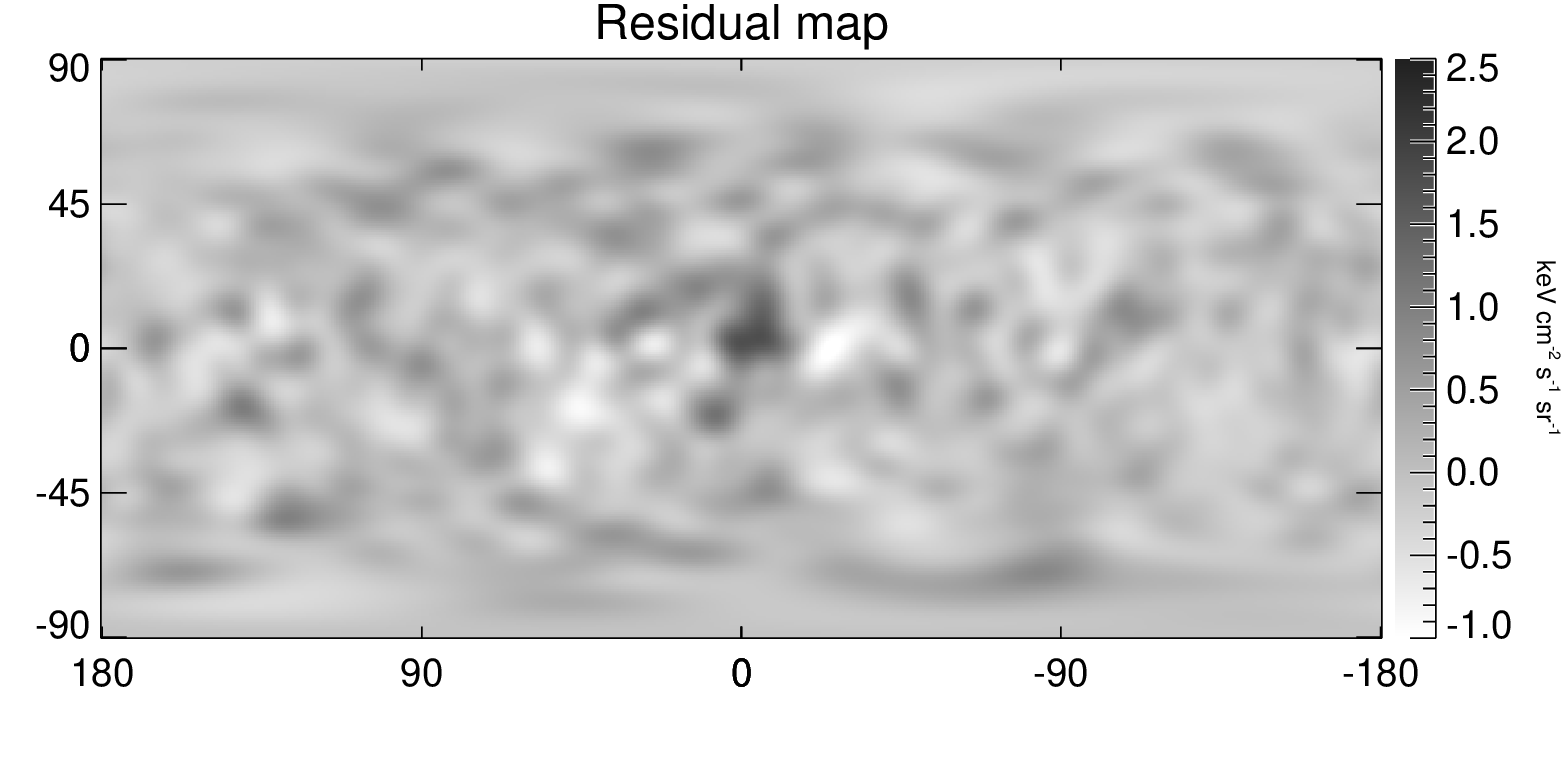}
        \includegraphics[width=0.47\textwidth]{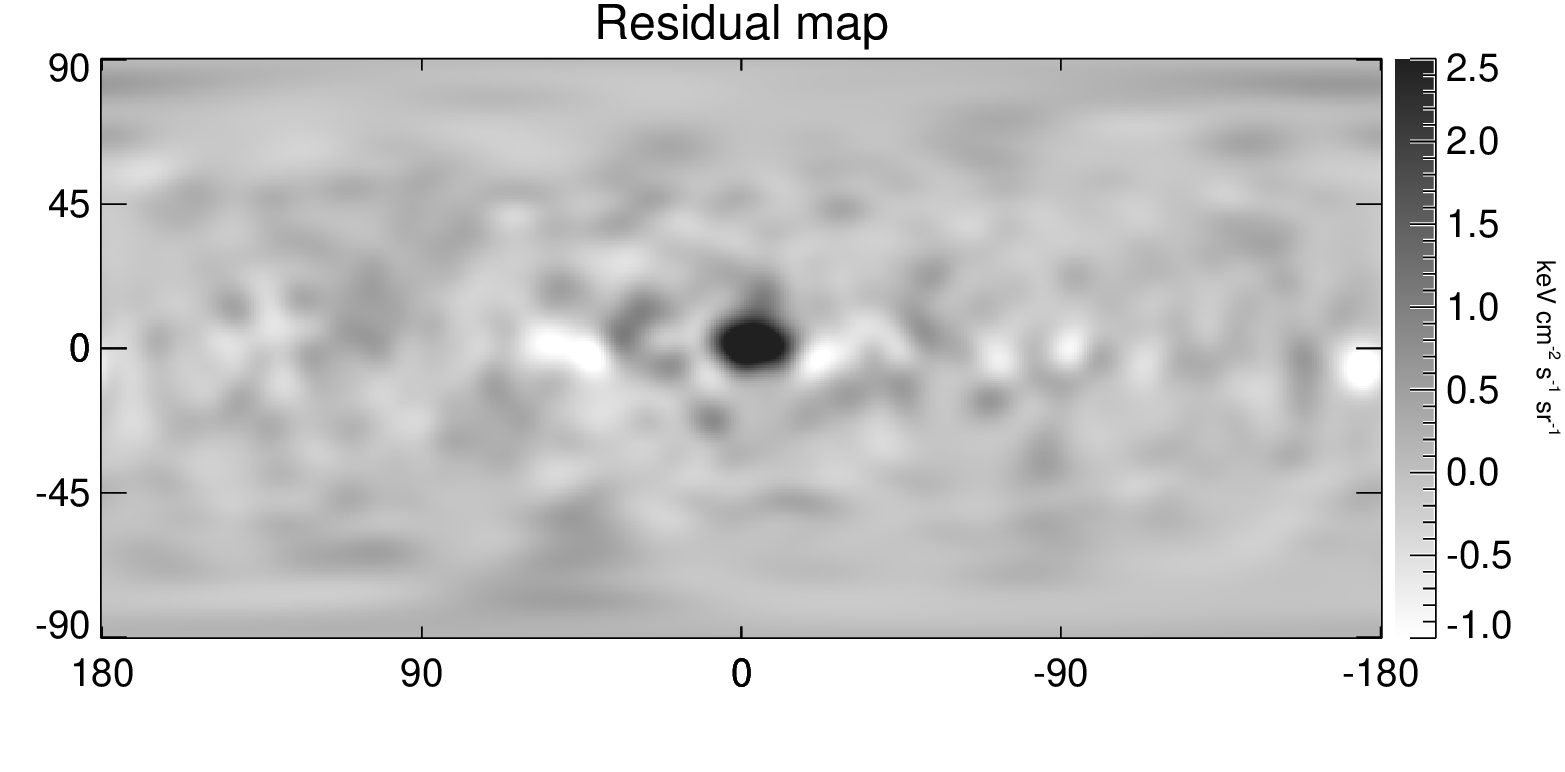}
        \includegraphics[width=0.47\textwidth]{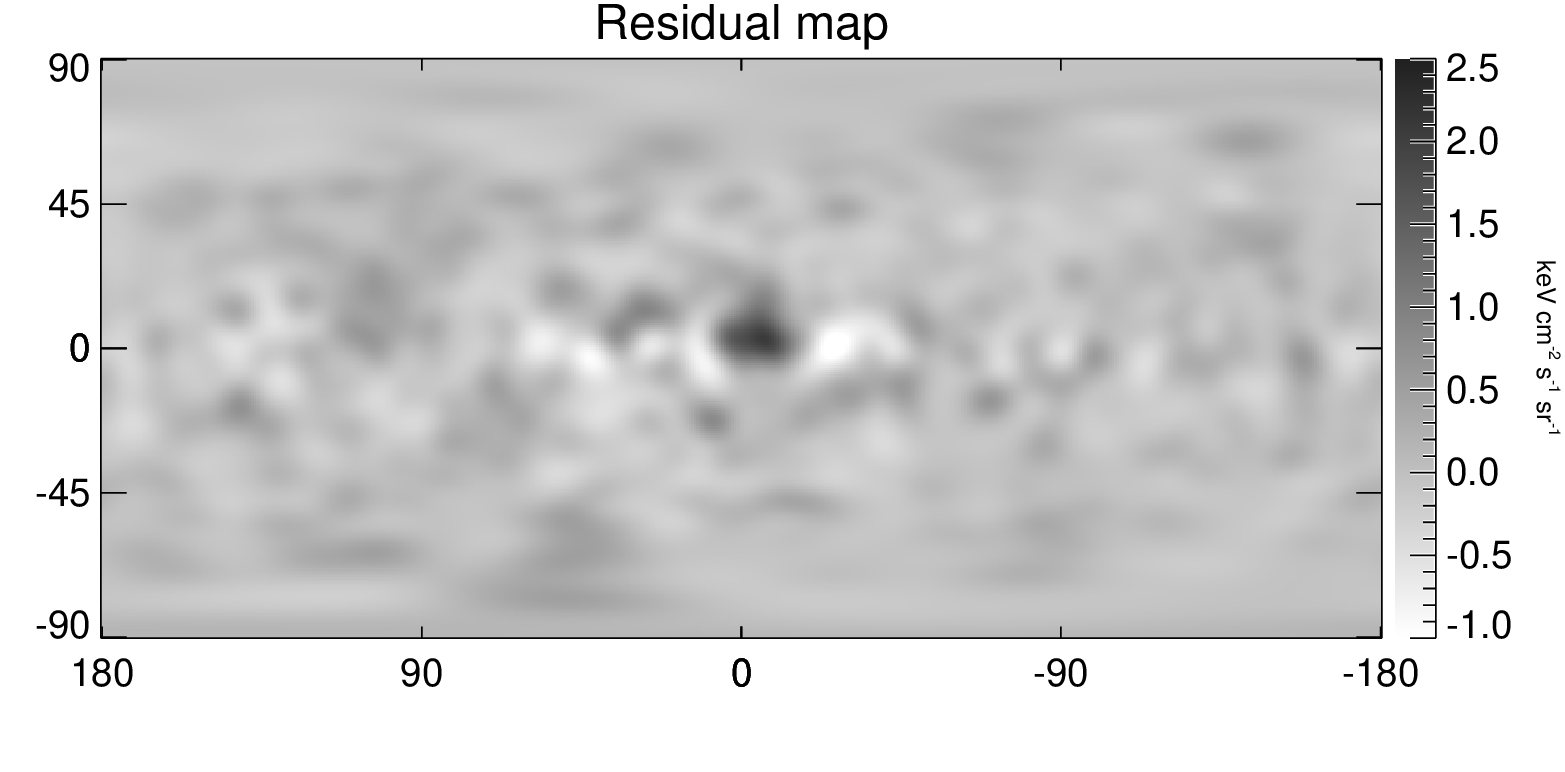}
    \end{center}
\caption{Residual maps the same as the \emph{lower right} panel of
Figures \ref{fig:fig3} and \ref{fig:fig4}, but using
\texttt{SOURCE} events before (\emph{upper left panel}) and
after (\emph{upper right panel}) point source subtraction, or using
\texttt{ULTRACLEAN} events before (\emph{lower left} panel) and
after (\emph{lower right} panel) point source subtraction. }
\label{fig:fig5}
\end{figure*}

\begin{figure}[ht]
    \begin{center}
        \includegraphics[width=0.47\textwidth]{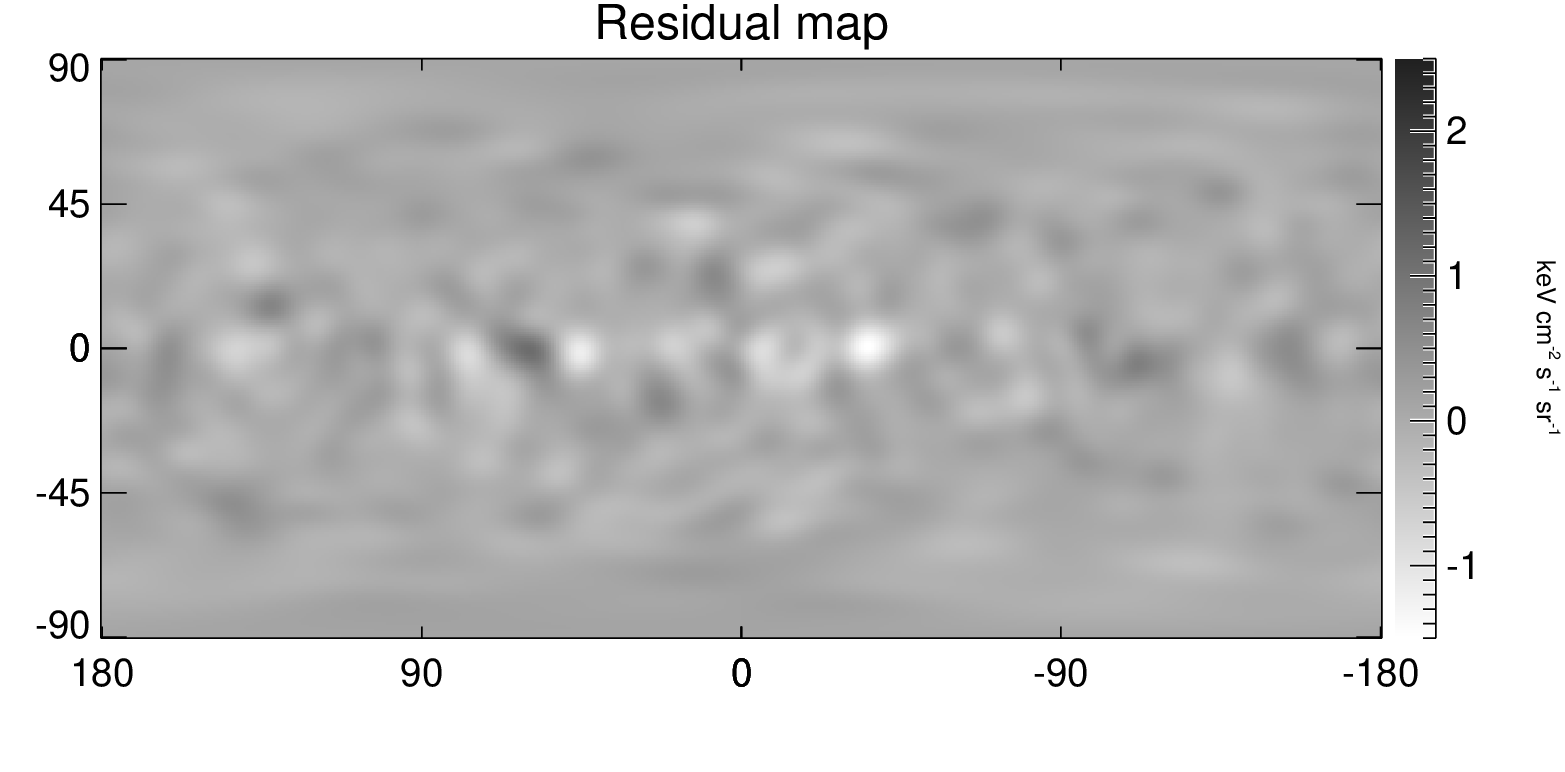}
        \includegraphics[width=0.47\textwidth]{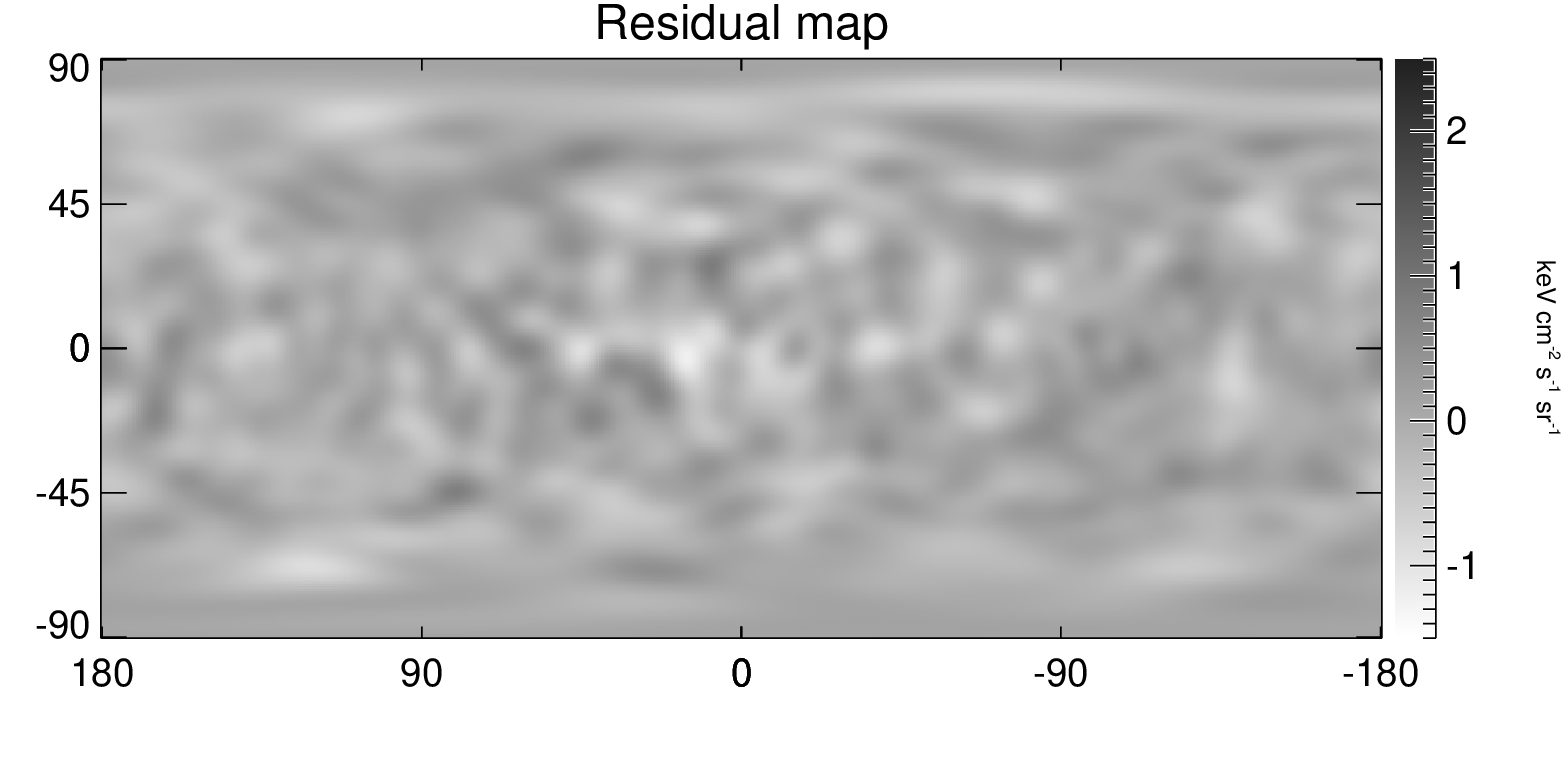}
    \end{center}
\caption{Residual maps by subtracting the average map of
$100-120$ GeV and $140-160$ GeV maps from the average map of
$80-100$ GeV and $160-180$ GeV maps (individual maps are
shown in \reffig{fig3}). Most of the large scale diffuse
gamma-ray structures (including the Galactic plane) and
bright point sources have been removed, and {\em no}
gamma-ray cuspy structure toward the Galactic center is
visible in the residual gamma-ray maps. The maps are
constructed using \Fermi-LAT 3.7 year \texttt{CLEAN} events
(\emph{upper panel}) and \texttt{SOURCE} events (\emph{lower
panel}), respectively. }
\label{fig:fig55}
\end{figure}

\begin{figure}[ht]
  \begin{center}
    \includegraphics[width=0.47\textwidth]{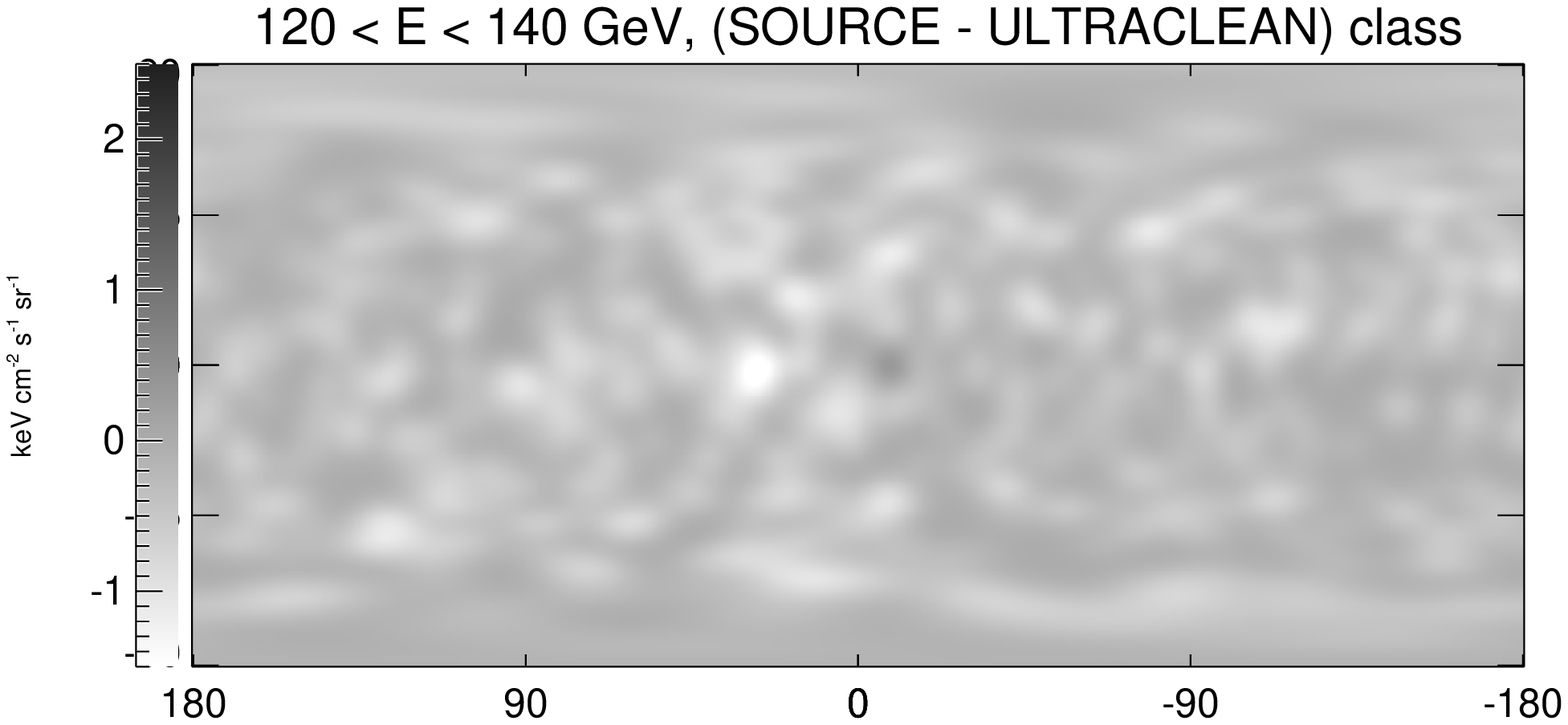}
  \end{center}
  \caption{Residual map after subtracting the $120-140$ GeV
    \texttt{ULTRACLEAN} map from the \texttt{SOURCE} map, with both smoothed
    to FWHM = $10\degree$. The map is consistent with Poisson noise, with no
    excess toward the Galactic center.}
  \label{fig:fig6}
\end{figure}

\begin{figure*}[ht]
    \begin{center}
        \includegraphics[width=0.8\textwidth]{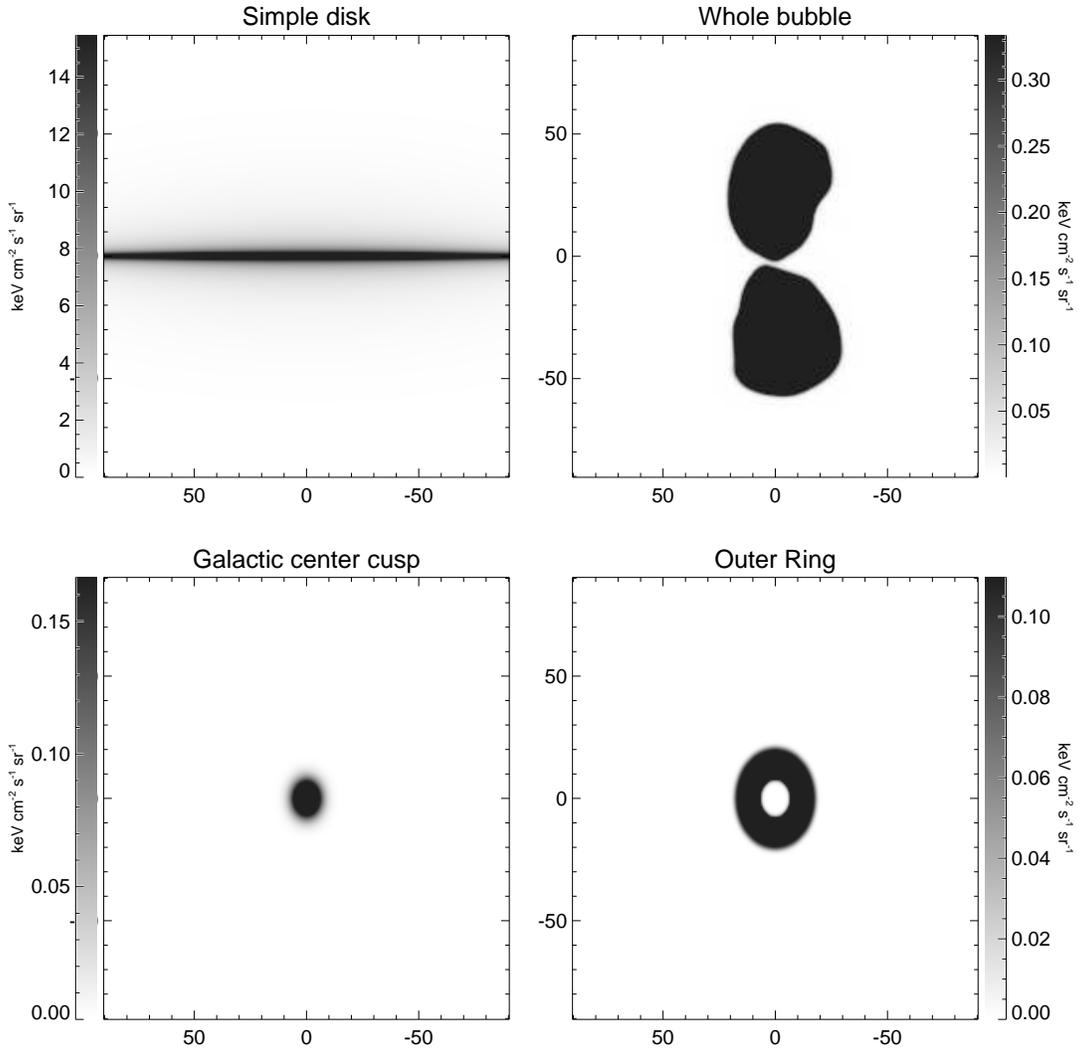}
    \end{center}
\caption{Spatial templates used in
the Poisson likelihood analysis. \emph{Upper left:}
Galactic disk template, \emph{upper right:} \Fermi\
bubble template, \emph{lower left:} gamma-ray
cuspy template as a Gaussian distribution with FWHM =
4$\degree$, \emph{lower right:} outer ring template as
a Gaussian distribution with FWHM = 10$\degree$, but masking
out the central region where the gamma-ray cuspy
template is. In \reffig{fig8} and \reffig{fig10}, we split
the \Fermi\ bubble template into two components one with
$|b| > 30\degree$ and the other with $|b| < 30\degree$. }
\label{fig:fig7}
\end{figure*}

\begin{figure}[ht]
    \begin{center}
	\includegraphics[width=0.45\textwidth]{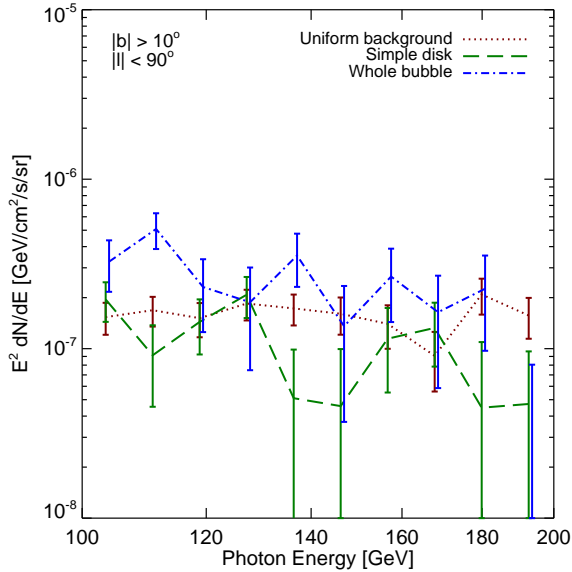}
    \end{center}
\caption{Spectral energy distribution of
three diffuse gamma-ray templates.  The disk-correlated
emission (\emph{green dashed}) approximately traces the
inverse Compton and bremsstrahlung components. The spectrum
of the uniform emission (\emph{dotted brown line}) includes
the isotropic part of the extragalactic background and
cosmic-ray contamination. We have not included the
dust-correlated emission or \emph{Loop I} template as in
\cite{FermiBubble}, because the expected spectrum is softer
and we do not expect a significant contribution from these
components at $E \gtrsim 100$ GeV.  Vertical bars show the
marginalized 68\% confidence range derived from the
parameter covariance matrix for the template coefficients in
each energy bin.  }
\label{fig:fig8}
\end{figure}

\begin{figure*}[ht]
    \begin{center}
	\includegraphics[width=0.45\textwidth]{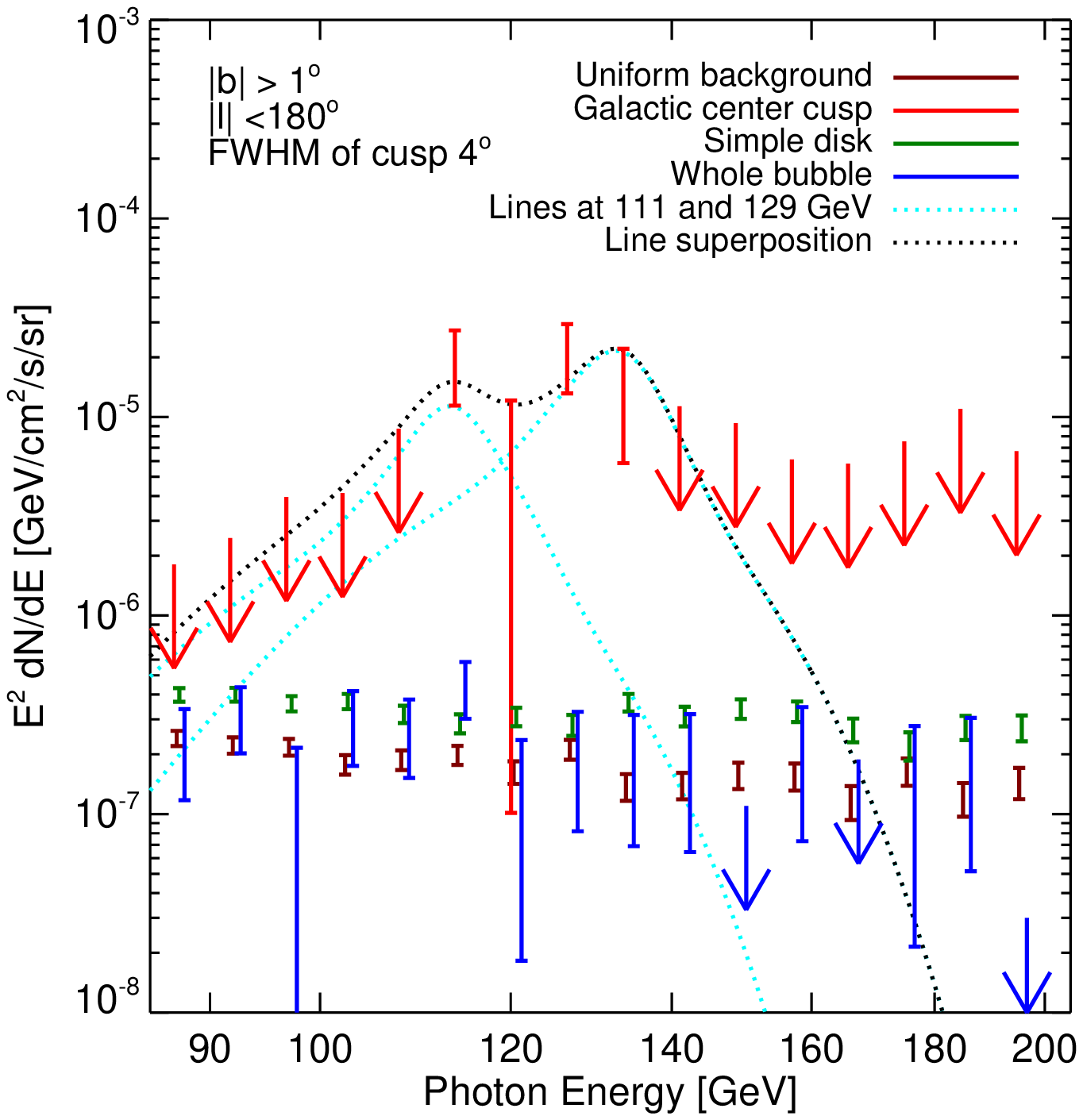}
        \includegraphics[width=0.45\textwidth]{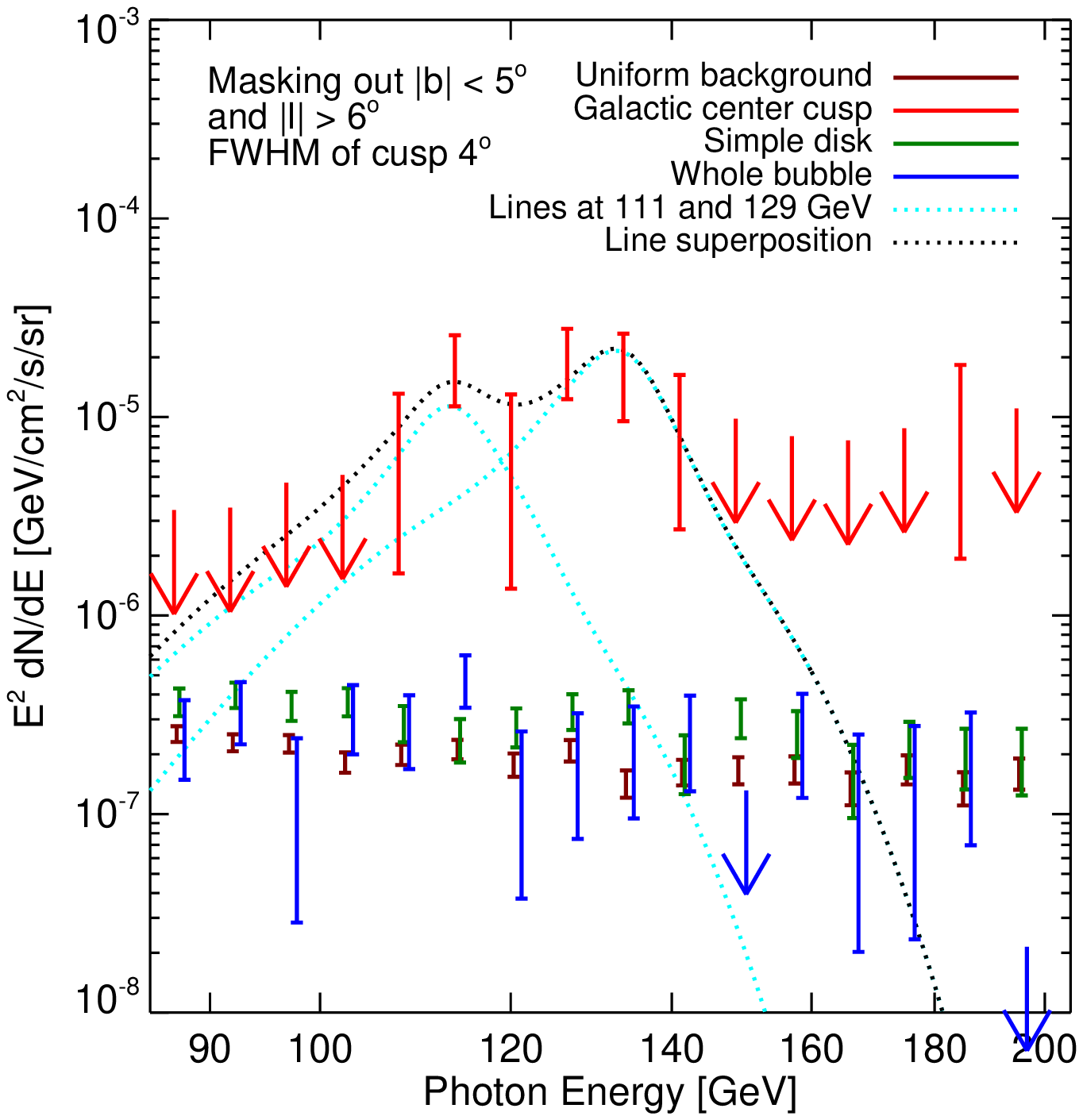}
    \end{center}
\caption{\emph{Left panel:} Spectral energy distributions of
the templates listed in the figure legend. In the \emph{left
panel}, we use \texttt{CLEAN} events with $|b|>1\degree$ and
all longitudes.  Besides the disk-correlated emission
(\emph{green}), uniform emission (\emph{brown}), and the 
\Fermi\ bubble template (\emph{blue}), the cusp component
modeled as a FWHM = 4$\degree$ Gaussian in the GC (\emph{red}) 
has been included.  Vertical bars show the marginalized 68\%
confidence range derived from the parameter covariance
matrix for the template coefficients in each energy bin. Arrows indicate $1
\sigma$ upper limits.   
For reference, we overplot lines centered at 111 GeV and 129 GeV
(\emph{dotted cyan}) convolved with a three-Gaussian approximation of the LAT
instrumental response~\citep{Edmonds:thesis}, and their
sum (\emph{dotted black}).  The line
centers and amplitudes are determined from a fit to the spectrum in the right
panel (see text). 
\emph{Right panel:} the same as the left panel but
using data masking out $|b|<5\degree$ and $|l|>6\degree$. 
}
\label{fig:fig9}
\end{figure*}


\begin{figure}[ht]
\begin{center}
\includegraphics[width=0.45\textwidth]{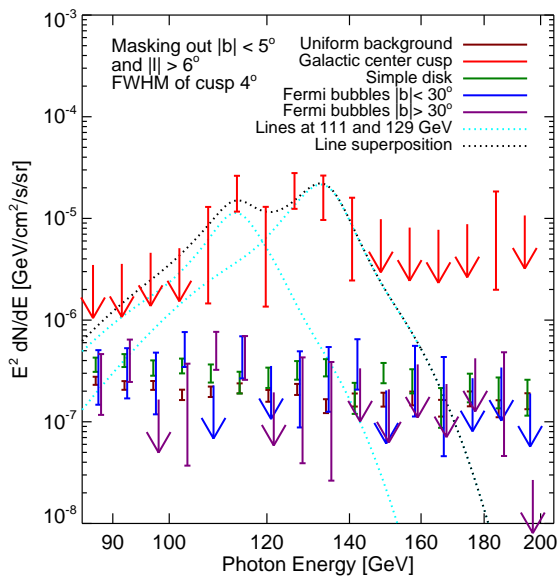}
\end{center}
\caption{Same as right panel of \reffig{fig9} but splitting the bubble template into two regions one with $|b|>30\degree$ and the other with $|b|<30\degree$. }
\label{fig:fig10}
\end{figure}


\begin{figure*}[ht]
    \begin{center}
	\includegraphics[width=0.45\textwidth]{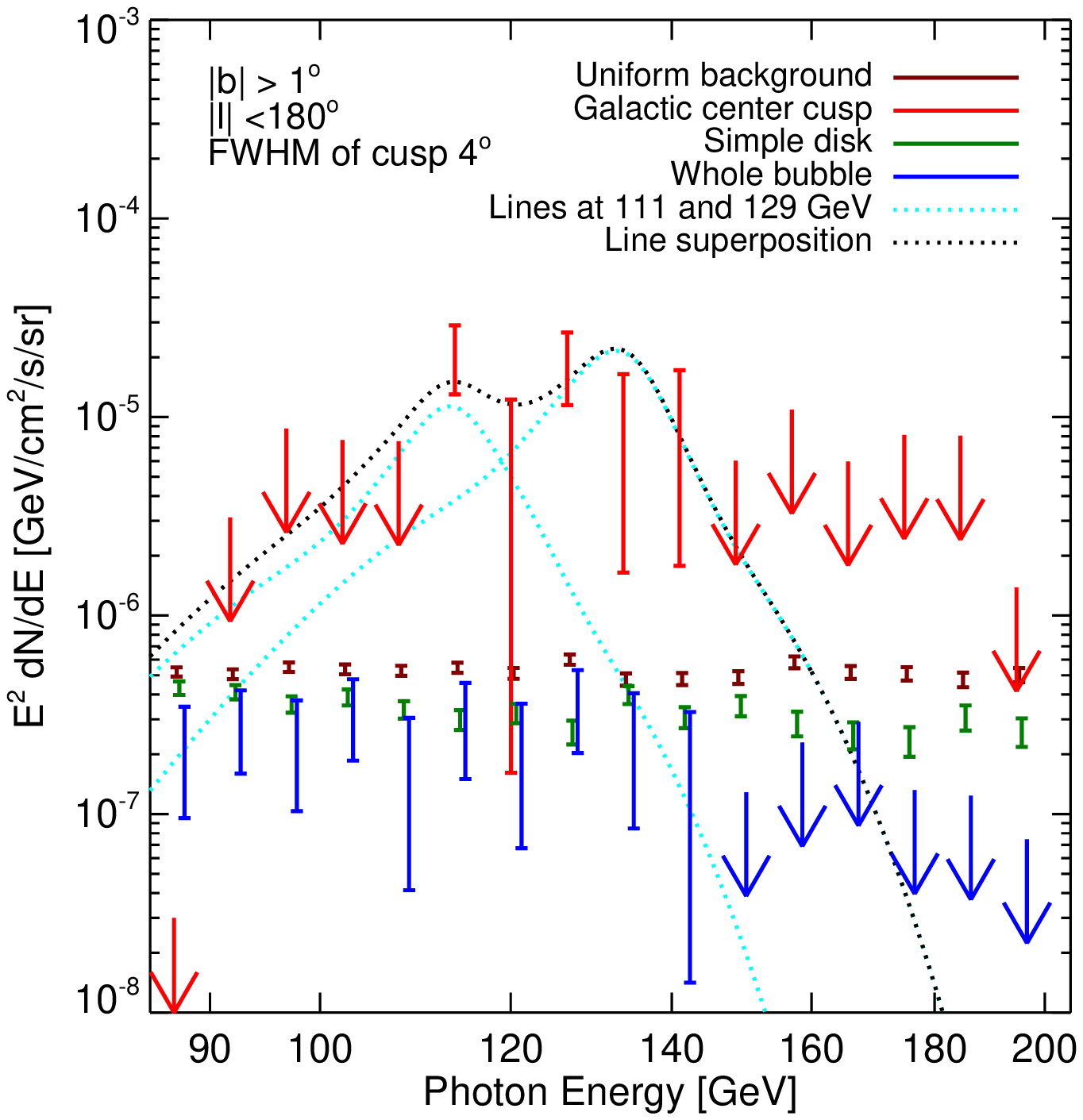}
        \includegraphics[width=0.45\textwidth]{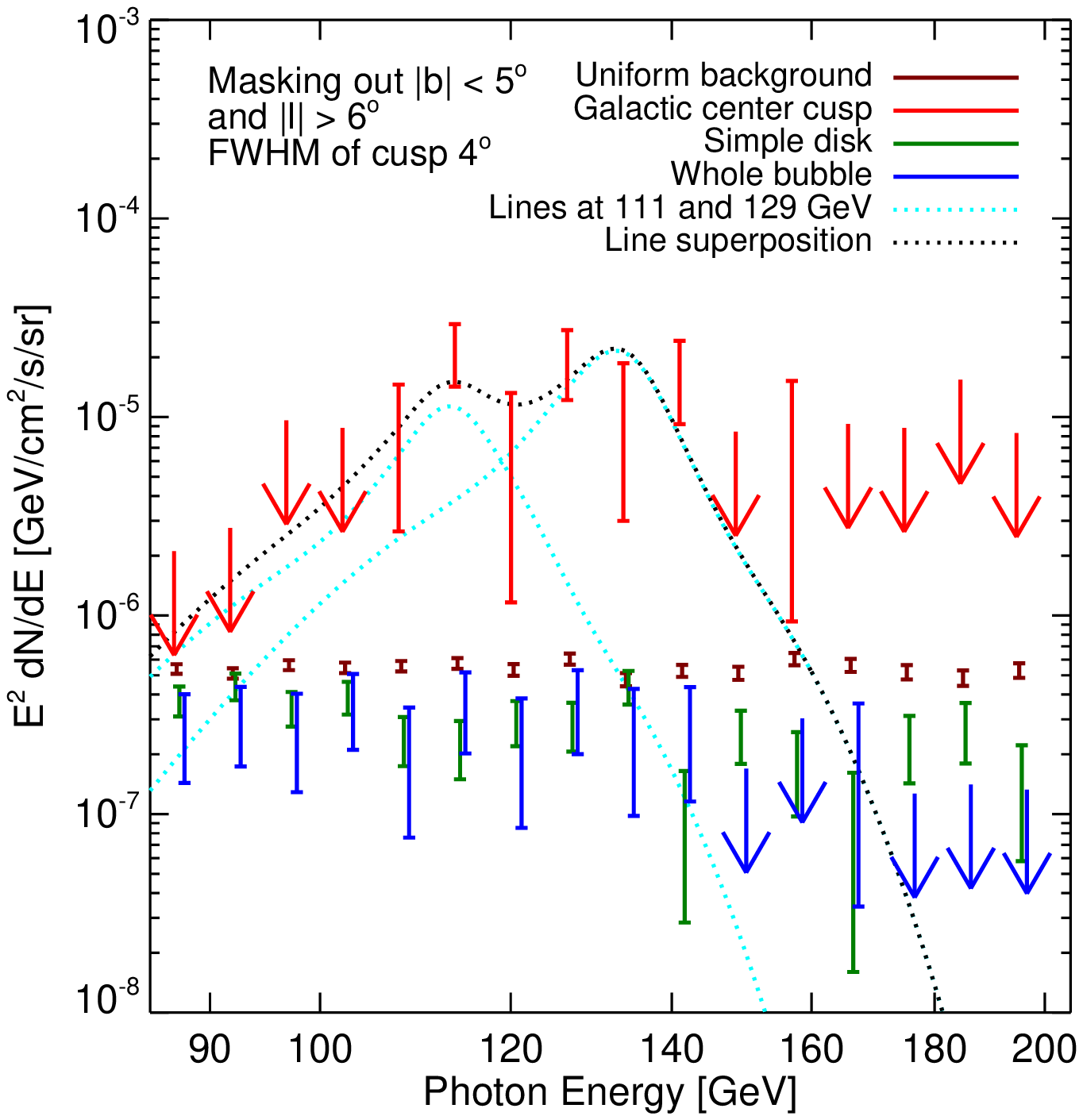}
    \end{center}
\caption{Same as \reffig{fig9} but using \texttt{SOURCE}
events instead of \texttt{CLEAN}. The level of the uniform
background is more than a factor of 2 higher than that shown
in \reffig{fig9}. However, the resulting energy spectrum of
the gamma-ray cusp (\emph{red dashed} line) is quite similar
to that shown in figures \ref{fig:fig9} and \ref{fig:fig10}.
}
\label{fig:fig12}
\end{figure*}

\begin{figure}[ht]
    \begin{center}
        \includegraphics[width=0.45\textwidth]{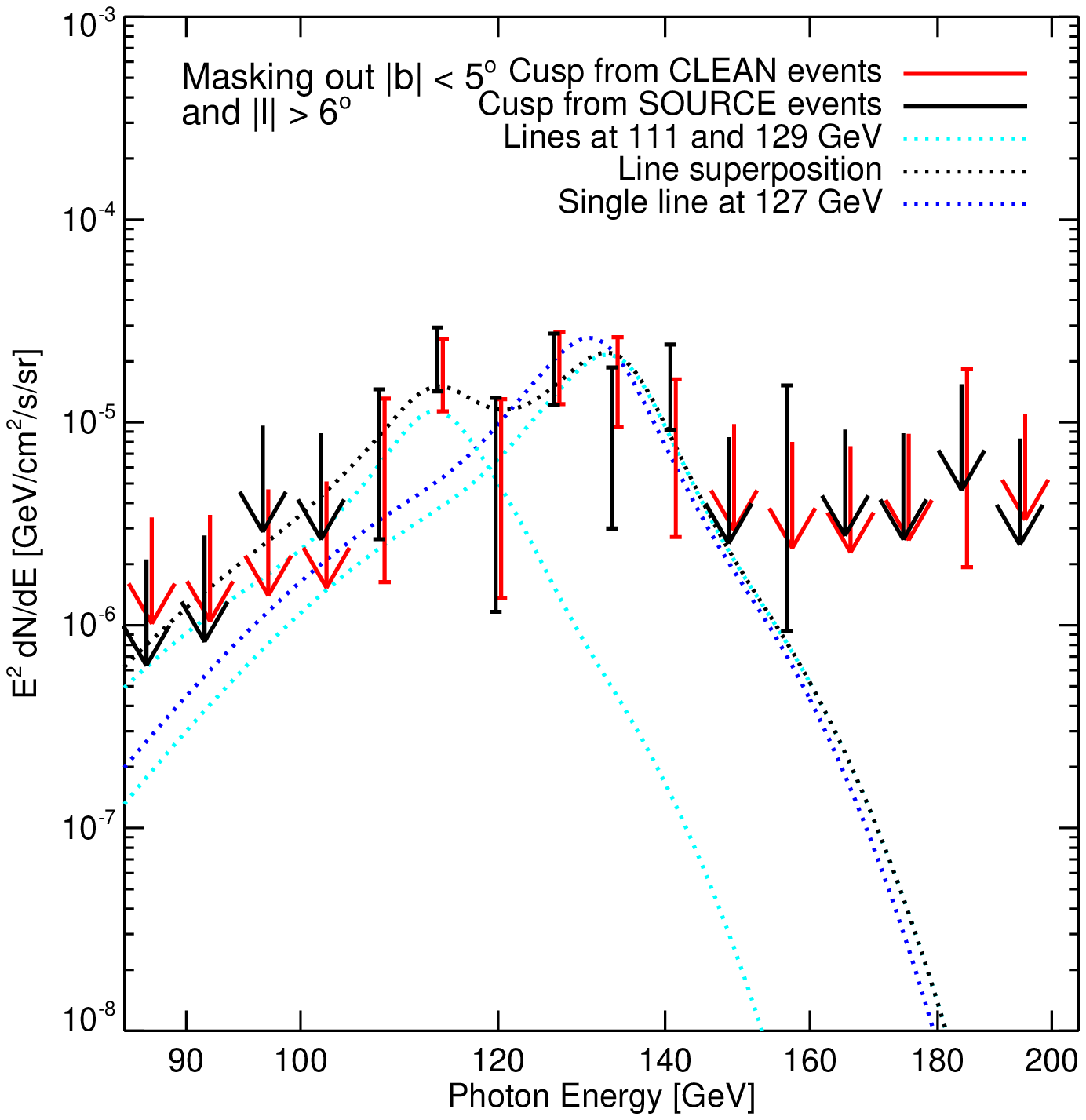}
    \end{center}
\caption{Same as right panel of \reffig{fig9} but overplot
the best fit one gamma-ray line profile convolved with the
instrument response, and compared with the best fit two-line
profile.}
\label{fig:fig13}
\end{figure}

The ``smoking-gun'' signal of annihilating dark matter would be a
monochromatic gamma-ray line (or lines) in a region of high dark matter
density, either in local dwarf galaxies or in the Galactic center.  This line
could be produced by dark matter decays or annihilations into two photons, or
two-body final states involving one photon plus a Higgs boson, Z boson, or
other chargeless non-SM particle.  
In most models, dark matter does not annihilate directly to photons,
but in models where it annihilates to charged lepton pairs, there may be loop
interactions that produce photons in two-body final states, yielding spectral
lines.  These lines can provide a signature of dark matter even in a
complicated astrophysical environment, because no known astrophysical process
produces gamma-ray lines at $E\gg1$ GeV.  However, these loop processes would
be suppressed by 1-4 orders of magnitude compared to the total annihilation or
decay rate~\citep[e.g.][]{Bergstrom:1997}. The unprecedented sensitivity and
spectral resolution of the Large Area Telescope \citep[LAT;][]{Gehrels:1999ri,
  Atwood:2009} aboard the \emph{Fermi Gamma-ray Space Telescope} make it
possible to search for dark matter annihilation lines over the whole sky up to
a few hundred GeV.

However, other signals can mimic a weak line.  Features such as a spectral
edge or a broken power law can be mistaken for a line in noisy data,
especially when smoothed by the instrumental response.  Thus careful
separation of possibly diffuse dark matter emission from other
diffuse components is crucial.

For example, the \fb\ extend $\sim$10 kpc above and below
the Galactic center (GC)~\citep{FermiBubble}. Their emission has a flat energy
spectrum in $E^{2}dN/dE$ from hundreds of MeV to $\sim$100 GeV implying the
ability of the \Fermi\ bubbles to accelerate cosmic ray (CR) electrons up to
$\sim$TeV, if the gamma rays are produced by inverse Compton (IC) scattering
of the Cosmic Microwave Background (CMB) photons~\citep{FermiBubble}. If there
is a spectral break in the \Fermi\ bubble spectrum at high energy, the
superposition of such a spectrum with other soft-spectrum diffuse
components might mimic a bump feature on top of a continuum power
law~\citep{Profumo:2012}.

In this work, we use 3.7 years data of LAT to study the
diffuse gamma-ray emission toward the inner Galaxy at $80 <
E_\gamma < 200$ GeV. In \Refsec{data} we describe our LAT
data selection and map making. In \Refsec{evidenceforhalo},
we show that the gamma-ray maps reveal a novel gamma-ray
cusp toward the Galactic center. We characterize the
morphology of the gamma-ray cusp and employ regression
template fitting to determine its energy spectrum in
\Refsec{cuspspec}. In \Refsec{profiles} we study the
detailed spatial profile of the cusp structure by fitting various templates.
Section \ref{sec:validation} contains analysis of the spectral
line profile,
and searches for instrumental effects in photon samples from the Galactic
plane and the Earth's limb.  We derive the energy spectrum of the cusp
assuming various dark matter density profiles in \Refsec{nfw}. 
We propose a modified survey strategy in \Refsec{strategy} that would allow
\Fermi-LAT to confirm this gamma-ray line signal in $\sim$1 year with no
trials factor, and we summarize our main findings in \Refsec{conclusion}.


\section{Map Construction}
\label{sec:data}

For this project, we constructed full-sky maps from the LAT event files
as in our previous work \citep{fermihaze, FermiBubble, FermiJet}, except that
we now use 3.7 years of Pass 7 (\texttt{P7\_V6}) data.  

\subsection{\emph{Fermi} data selection}
\label{para:fermidata}

The \Fermi\ LAT is a pair-conversion telescope, in which incoming photons
convert to $e^+e^-$ pairs, which are then tracked through the detector.  The
arrival direction and energy of each event are reconstructed, and the time of
arrival recorded.  Event files for every week of the mission are available on
the Internet, and it is from these files that we build our maps. 

The point spread
function (PSF) is about 0.8$\degree$ for 68\% containment at
1 GeV and decreases with energy as $r_{68}\sim E^{-0.8}$,
asymptoting to $\sim$ 0.2$\degree$ at high energy. The LAT is
designed to survey the gamma-ray sky in the energy range
from about 20 MeV to several hundreds of GeV.

We use the latest publicly available data and instrument response functions,
known as Pass 7 (\texttt{P7\_V6})\footnote{Details at
\texttt{http://fermi.gsfc.nasa.gov/ssc/data/analysis/}
\texttt{documentation/Pass7\_usage.html}}. For most figures in
this work we use the \texttt{CLEAN} event class, which has larger effective
area than \texttt{ULTRACLEAN} and lower background than \texttt{SOURCE}.
In a few cases, we show figures made with \texttt{ULTRACLEAN} or
\texttt{SOURCE} events as evidence that this choice has no
qualitative effect on our results.

Photons coming from the bright limb at Earth's horizon, dominantly produced by
grazing-incidence CR showers in the atmosphere, are a potential
source of contamination. We minimize this background by selecting events
with zenith angle less than $100\degree$ as suggested in the \Fermi\
Cicerone\footnote{\texttt{http://fermi.gsfc.nasa.gov/ssc/data/analysis/documentation/}.}.
We also exclude some time intervals, primarily while \Fermi\ passes
through the South Atlantic Anomaly.

\subsection{Map making}
\label{para:mapmaking}

We generate full-sky maps of counts and exposure using
HEALPix, a convenient equal-area iso-latitude full-sky
pixelization widely used in the CMB
community.\footnote{HEALPix software and documentation can
be found at \texttt{http://healpix.jpl.nasa.gov}, and the
IDL routines used in this analysis are available as part of
the IDLUTILS product at
\texttt{http://sdss3data.lbl.gov/software/idlutils}.}
Spherical harmonic smoothing is straightforward in this pixelization, and we
smooth each map by the kernel required to obtain an approximately Gaussian PSF
of some target FWHM, usually $10\degree$. We generate maps for
front- and back-converting events separately, smooth them to a common PSF, and
then combine them.

We construct maps both with and without point source subtraction.  We subtract
point sources listed in the Second
\Fermi-LAT catalog (2FGL), which is based on 24 months of P7\_V6 LAT
observations.\footnote{\texttt{http://fermi.gsfc.nasa.gov/ssc/data/access/lat/2yr\_catalog},
the file we used is \texttt{gll\_psc\_v07.fit}}
The PSF and effective area of the \Fermi-LAT varies with energy, and we
subtract each point source from the maps in each energy bin, using the
in-flight version of the PSF contained in the P7\_V6 IRFs. 

For the 400
brightest and 400 most variable sources, the subtraction is
noticeably imperfect at lower energies (and we assume it is also at the higher
energies used in this work), so we interpolate over the core of the
PSF after subtracting the best estimate. We also mask out sources
including Geminga, 3C 454.3, and LAT PSR J1836+5925 and
large sources like Orion and the Magellanic Clouds as we did
in the previous papers~\citep{FermiBubble, FermiJet},
although they are unlikely to be a problem at $E >
80$ GeV, where we are searching for lines.

We produce the exposure maps using the \texttt{gtltcube} and
\texttt{gtexpcube2} tasks in the Fermi Science Tools.  For bright sources, the
exposure is set to zero for excised pixels.  For the smoothed maps, both the
count map and exposure map are smoothed, and then divided.  At high energies,
where the PSF is small, this effectively interpolates over the masked pixels.

\section{A model-independent search for line-emitting regions}
\label{sec:evidenceforhalo}
We begin by constructing maps in broad (20 GeV) energy bins and taking linear
combinations that cancel most Galactic emission.  This has the potential to
reveal a component with an unusual spectrum in a model-independent way. 

In Figures \ref{fig:fig1} and \ref{fig:fig2}, we show the full sky map in four
energy bins in the range 100 to 180 GeV, using
\texttt{CLEAN} and 
\texttt{SOURCE} events, respectively. Even after nearly
four years of observation, the number of gamma-ray
photons with $E \gtrsim 100$ GeV is still quite limited and
the maps are Poisson noise dominated. In order to inspect
diffuse gamma-ray structure, we smooth the maps with 
a Gaussian kernel of FWHM = $10\degree$.  Smoothed \texttt{CLEAN} maps in
various energy bins with $E > 80$ GeV, before and after
subtracting point sources, are shown in Figures \ref{fig:fig3} and 
\ref{fig:fig4}, respectively. To look for any diffuse
gamma-ray emission component with a distinctive energy
spectrum, we examine various linear combinations of maps that cancel out the
Galactic plane emission and visually inspect the residual maps.

Interestingly, when we subtract the average map (averaged in 
$E^2 dN/dE$ units) of the $80-100$
GeV, $100-120$ GeV, $140-160$ GeV, and $160-180$ GeV maps
from the $120-140$ GeV map, most of the large scale diffuse
gamma-ray structures (including the relatively bright
Galactic plane) and visible point sources have been largely
removed. However, a resolved cuspy structure toward the inner
Galaxy with a size slightly larger than the smoothing kernel
is revealed
as the only visible structure in the residual gamma-ray
map. We have repeated the analysis only subtracting $80-100$
GeV and $100-120$ GeV maps from the $120-140$ GeV map, or
only subtracting $140-160$ GeV and $160-180$ GeV maps, and
we obtained similar residual structure toward the Galactic
center. We
also repeat this exercise using \texttt{SOURCE} and
\texttt{ULTRACLEAN} event classes and the residual maps are
shown in \reffig{fig5}.  The GC excess is similar in each case,
and the rest of the
gamma-ray sky is consistent with Poisson noise. This excess
at $\sim 120-140$ GeV strongly suggests a novel diffuse
gamma-ray component toward the Galactic center with unusual
spectrum. Furthermore, this energy range coincides with the
recently suggested tentative signature of gamma-ray excess
at 130 GeV~\citep{Bringmann:2012,Weniger:2012}, which is
under active debate in the literature
\citep{Boyarsky:2012,Tempel:2012,Profumo:2012}. Assuming the
distance to the Galactic center is $R_\odot$ = 8.5 kpc, the
size of the gamma-ray cusp is
$\lesssim 1$ kpc.

We show in \reffig{fig55} the difference map between the
average map of $80-100$ GeV and $160-180$ GeV maps and the
average map of $100-120$ GeV and $140-160$ GeV maps. The
difference map is consistent with Poisson noise and {\em no}
diffuse gamma-ray excess toward the inner Galaxy is
visible. In order to test whether the excess
is due to residual cosmic
ray contamination, we subtract \texttt{ULTRACLEAN} sky maps
from \texttt{SOURCE} sky maps. This residual map should be
mostly dominated by cosmic rays since a large fraction of
the real gamma-ray photons have been removed. Indeed,
\reffig{fig6} demonstrates that there is no excess toward
the inner Galaxy in this map, thus we can rule out the
possibility that the central excess is due to cosmic
ray contamination in the LAT data. 

Toward the inner Galaxy, the \Fermi\ bubbles extend $\sim
50\degree$ above and below the Galactic center, with a width
of $\sim 40\degree$ in longitude. The gamma-ray emission
associated with these bubbles has a significantly harder
spectrum ($dN/dE \sim E^{-2}$) than the inverse Compton
emission from electrons in the Galactic disk, or the
gamma-rays produced by decay of $\pi^0$ from proton-ISM
collisions. We note that the morphology of the resolved
gamma-ray cusp has a different shape from the bubbles, and
the bubble structure has been largely cancelled out and not
visible in the residual maps (as shown in
e.g. \reffig{fig5}). Thus we conclude that the new
gamma-ray cusp has no obvious connection with the
bubbles. We will also demonstrate this more explicitly in
\Refsec{cuspspec}.

\begin{figure}[ht]
\begin{center}
\includegraphics[width=0.45\textwidth]{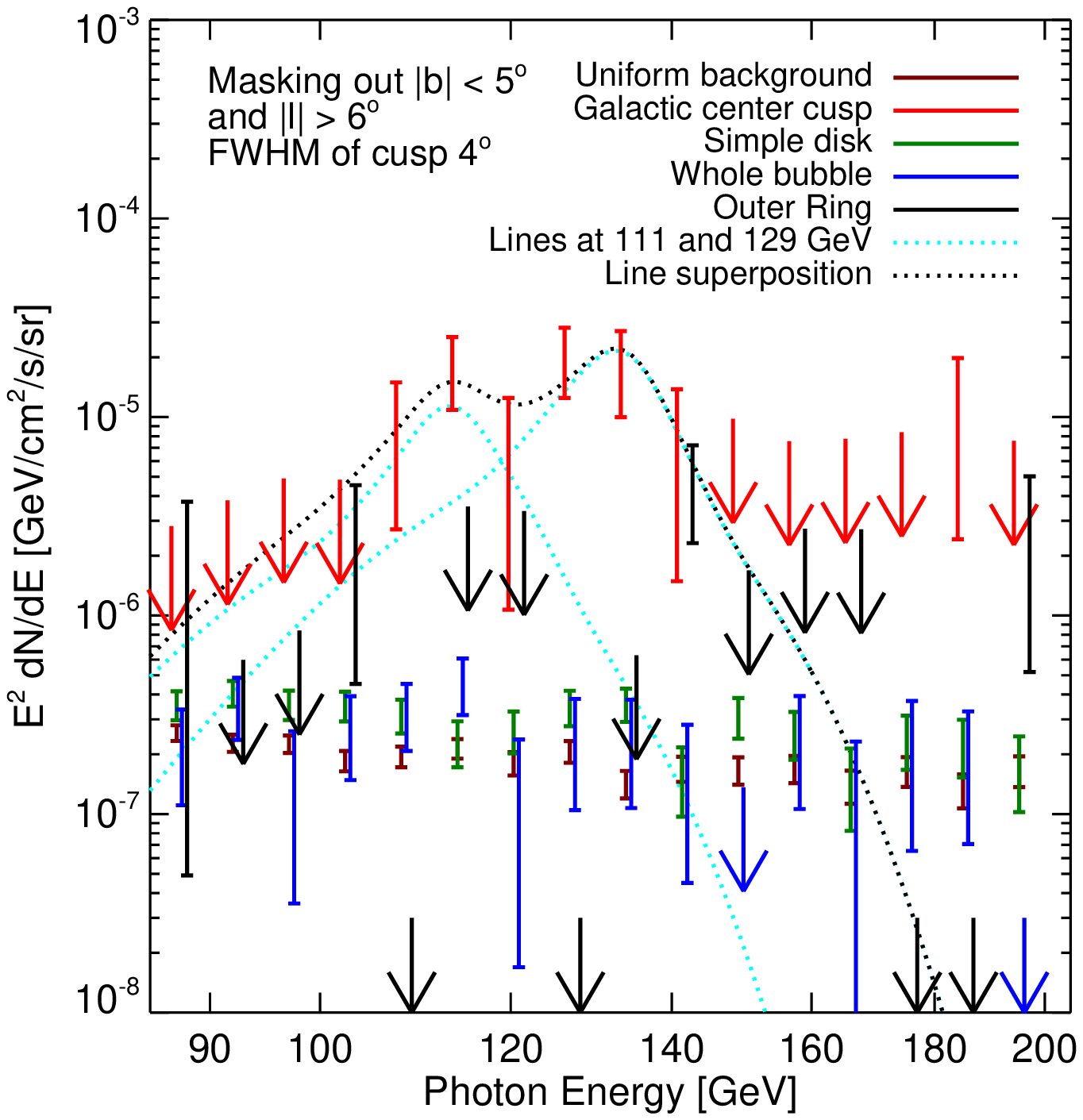}
\end{center}
\caption{Same as right panel of \reffig{fig9} but including extra outer ring template. }
\label{fig:fig14}
\end{figure}

\section{Energy Spectrum of the Gamma-ray Cusp}
\label{sec:cuspspec}

In \refsec{evidenceforhalo} we argued for the existence of an excess with a concentrated cuspy shape in the GC
at $E=120-140$ GeV. 
This motivates a more careful investigation of the energy spectrum and spatial
distribution of the emission. 
\subsection{Template regression}

As in our previous work \citep[e.g.,][]{FermiBubble}, we consider linear
combinations of spatial templates and compute the likelihood
that the gamma-ray maps are described by a given linear
combination.  We fit a coefficient for each emission
component using a multi-linear regression of simple
templates, one energy bin at a time.  This technique provides an estimate of
the energy spectrum of each component using as few physical assumptions as
possible. 

By combining results from 16 logarithmically-spaced energy
bands from 85 GeV to 200 GeV, we determine the spectral
energy distribution for each component. In each fit, we
model the ``conventional'' emission using three simple
templates: a Galactic disk model, the \Fermi\ bubbles, and a
uniform background.  The Galactic disk template has the
functional form $(\csc|b|)-1$ in latitude and is a Gaussian
($\sigma_\ell=80\degree$) in longitude, as in
\cite{FermiBubble}.  The disk model mostly accounts for
gamma rays from the Galactic plane including those produced
by point sources, ISM emission ($\pi^0$ and Bremsstrahlung)
and inverse Compton scattering.  The uniform background
template absorbs the isotropic background due to
extra-galactic emission and misclassified charged particle
contamination, including heavy nuclei at high energies. We
show maps of these templates in \reffig{fig7}.

For each set of model parameters, we compute the Poisson log
likelihood, \be \ln {\mathcal L} = \sum_i k_i\ln\mu_i -
\mu_i - \ln(k_i!), \ee where $\mu_i$ is the model counts
map (i.e., linear combination of templates times exposure
map) at pixel $i$, and $k$ is the map of observed
counts. The last term is a function only of the observed
maps.  We compute parameter errors in the Gaussian approximation by
inverting the matrix of second partial derivatives of $-\ln
{\mathcal L}$ to obtain the covariance matrix, and taking
the square root of the diagonals.  The $1\sigma$ Gaussian
error corresponds to $\Delta\ln {\mathcal L} = 1/2$.

Template-correlated spectra for the 3-template fit are shown
in \reffig{fig8}. The energy spectra show no significant deviation from a
power law for any of the three components. 
This fact, together with the distinct spatial morphology of the
gamma-ray cusp suggests the need to include a cusp template (shown in
\reffig{fig7}) in the model. 
Inclusion of this template barely alters the derived spectrum of the first three
components, but yields significant coefficients for the cusp 
from 110 to 140 GeV (\reffig{fig9} and \reftbl{dmfits}). 
The fact that the bubble coefficients show no such bump indicates that the
bubble structure is unrelated to the $\sim130$ GeV excess. 
We find that the surface brightness of the center of the 
cusp is nearly two orders of magnitude greater than
that of the \Fermi\ bubble structure, but only over a limited energy range. 

In \reffig{fig9}, we have done the fit with two different
spatial masks: one excludes $|b| < 1\degree$ to avoid
contamination from the Galactic disk close to the plane and
the Galactic center (including the Galactic ``ridge'');
another one excludes much more of the Galactic disk with
$|b| < 5\degree$ and $|l| > 6\degree$ without masking the
Galactic center region. We have obtained similar results for
both data cuts. The cusp structure emits gamma rays with a
luminosity of $(3.2\pm0.6)\times 10^{35}$ erg/s or
$(1.7\pm0.4)\times 10^{36}$ photons/sec.  The null
hypothesis of zero intensity of the cusp component is ruled
out by $5.0\sigma$. Given the energy resolution of the
\Fermi-LAT at $E \gtrsim 100$ GeV, the spectral excess at
$110 \lesssim E \lesssim 140$ GeV is consistent with
emission from one or two lines after considering the
line-spread function (LSF)~\citep{Edmonds:thesis}, which
strongly suggests the novel nature of the gamma-ray cusp as
no known astrophysical process can produce this feature.
Except for unexpected instrumental systematics or an
increasingly unlikely statistical fluke, a dark matter
annihilation signal from the inner Galaxy is the most likely
explanation.
In another variant of the fit, we split the bubble template
into two independent components in the fitting, high
latitude ($|b| > 30\degree$) and low latitude ($|b| <
30\degree$). The purpose is to demonstrate that the low
latitude bubble is also independent from the gamma-ray cusp.
Again, we find no sign of a bump in the spectra of other
diffuse gamma-ray components, but the cusp has a spectrum
with an excess at $110 - 140$ GeV and is consistent with
zero in the other bins (\reffig{fig10}).  Instead of using
\texttt{CLEAN} class, we have tried using \texttt{SOURCE}
class for the likelihood analysis, and obtained similar
results (\reffig{fig12}).

 \begin{table}
 \begin{center}
 \begin{tabular}{@{}rrrrr}
\hline
\hline
E range (GeV) & Energy & cusp (\texttt{CLEAN}) & cusp (\texttt{SOURCE})\\
\hline
$   84.9 -   89.5$
 &    87.2  &  -1.01 $\pm$ 4.42 & -2.19$\pm$ 4.30\\
$   89.5 -   94.5$
 &    92.0  &  -0.79 $\pm$ 4.28 & -1.53$\pm$4.29 \\
$   94.5 -   99.7$
 &    97.1  &  0.03 $\pm$ 4.64 & 4.37$\pm$5.26 \\
$   99.7 -  105.2$
 &   102.4  &  0.06 $\pm$ 5.04 & 3.05$\pm$5.77 \\
$  105.2 -  111.0$
 &   108.1  &  7.37 $\pm$ 5.73 & 8.61$\pm$5.95 \\
$  111.0 -  117.1$
 &  114.0   &  18.58 $\pm$ 7.25 & 21.80$\pm$7.57 \\
$   117.1 -   123.6$
 &    120.3 &  7.18 $\pm$ 5.82 & 7.19$\pm$6.03 \\
$   123.6 -   130.4$
 &    127.0 &  20.06 $\pm$ 7.75 & 19.78$\pm$7.61  \\
$   130.4 -   137.6$
 &    134.0 &  17.91 $\pm$ 8.38 & 10.82$\pm$7.83 \\
$   137.6 -   145.2$
 &    141.4 &  9.50 $\pm$ 6.78 & 16.71$\pm$7.50 \\
$   145.2 -   153.2$
 &    149.2 &  4.07 $\pm$ 5.73 & 3.07$\pm$ 5.36\\
$   153.2 -   161.7$
 &    157.4 &  1.70 $\pm$ 6.29 & 8.07$\pm$ 7.14\\
$   161.7 -   170.6$
 &    166.1 &  3.11 $\pm$ 4.50 & 4.34$\pm$ 4.88\\
$   170.6 -   180.1$
 &    175.2 &  3.08 $\pm$ 5.69 & 2.91$\pm$ 5.90\\
$   180.1 -   190.0$
 &    185.0 &  10.11 $\pm$ 8.18 & 7.07$\pm$ 8.34\\
$   190.0 -   200.5$
 &    195.2 &  3.99 $\pm$ 7.04 & 1.84$\pm$ 6.46\\
\hline
 \end{tabular}
 \end{center}
\caption{The template fitting coefficients and errors of the
  diffuse gamma-ray cusp correspond to the right panel of
  \reffig{fig9} and right panel of \reffig{fig12}. The
  gamma-ray luminosity in each energy range is shown in the
  unit of $\kevflux$.}
\label{tbl:dmfits}
\end{table}

The energy spectrum of the cusp is consistent with a single
spectral line (at energy $127.0\pm 2.0$ GeV with
$\chi^2=4.48$ for 4 d.o.f.).  But a pair of lines at $110.8\pm
4.4$ GeV and $128.8\pm 2.7$ GeV provides a marginally better
fit (with $\chi^2=1.25$ for 2 d.o.f.). We have compared the
best fit one line and two line profile with the measured
energy spectrum in \reffig{fig13}. The observation is
compatible with a $140.8\pm 2.8$ GeV WIMP annihilating
through $\gamma Z$ and $\gamma h$ assuming $m_h=125$ GeV
(with $\chi^2=3.33$ for 3 d.o.f.) or a $127.3 \pm 2.7$ GeV
WIMP annihilating through $\gamma \gamma$ and $\gamma Z$ (with
$\chi^2=1.67$ for 3 d.o.f.)~\citep[e.g.,][]{Weiner:2012}. 

The gamma-ray cusp appears to possess a symmetric
distribution around the Galactic center. To investigate
whether there is any more extended cusp component
contributing the excess at $120-140$ GeV, we include an extra ``outer ring''
template as
shown in \reffig{fig7}. The outer ring template is a FWHM=$10\degree$ Gaussian
with an $8\degree$ radius hole in the center.  Even with
this freedom, there is no significant change in the cusp spectrum
(\reffig{fig14}).  There was no
significant improvement of the likelihood for this model, and the spectrum of
the outer ring is consistent with zero. Our conclusion is that the gamma-ray
cusp is a
distinct component, and is centrally concentrated. 

\subsection{Trials factor}
\label{sec:trials}
We use a trials factor of 300 for the single-line fits centered on the Galactic center, and
6000 for fits that are off center.   This choice is based
on the fact that the LAT energy resolution is $\sim 10$\%
over most of the energy range, and a line anywhere from 1 to
300 GeV would have been just as impressive, yielding 60 energy bins.
Furthermore, a
broader line (or two lines near each other) has an additional
trials factor.  We allow an extra factor of 5, giving us
300.

For fits that recenter the cusp in $\ell$, we would have considered any center
with $\ell|<5\degree$ interesting, and the centering must change by $\pm
0.25\degree$ to make unit change in $TS$, so we take there to be 20
interesting spatial bins.  For single-line, off-center fits we use a trials
factor of 6000. 

The local significance of the centered Gaussian template is 5.0$\sigma$,
obtained by summing the significance of each bin in
quadrature (\reftbl{dmfits}).  After diluting the $p$ value
corresponding to $5.0\sigma$ by a factor of 300, we obtain a
significance corresponding to $3.7\sigma$.  We note that, if
the line is real, an additional 40\% more data will be
enough to obtain a $5.0\sigma$ detection, even with the
trials factor of 300. 

The previous trials factor is for a line \emph{anywhere}
with a range of widths.  On the other hand, if one asserts
that the line pair is from $\gamma\gamma$ and $\gamma$Z dark
matter annihilation channels, the trials factor can be
calculated as follows:

The two $\gamma$ lines could have been anywhere between the
Z mass and 300 GeV.  There are 12 log-spaced bins of width
10\% in that range. Also, there is no additional trials
factor for a width; we simply have two lines, and we let
their two amplitudes float as a free parameter, as well as
the WIMP mass. There is also a factor from the number of
line-producing scenarios: we could have seen a single line
or two lines from $\gamma\gamma$ and $\gamma$Z or $\gamma$Z
and $\gamma$h.  Considering these 3 scenarios, 
we assign a trials factor of 36 for the single-line fits. 
For off-center fits, we use $36\times 20=720$. 


\begin{figure*}[ht]
    \begin{center}
        \includegraphics[width=0.36\textwidth]{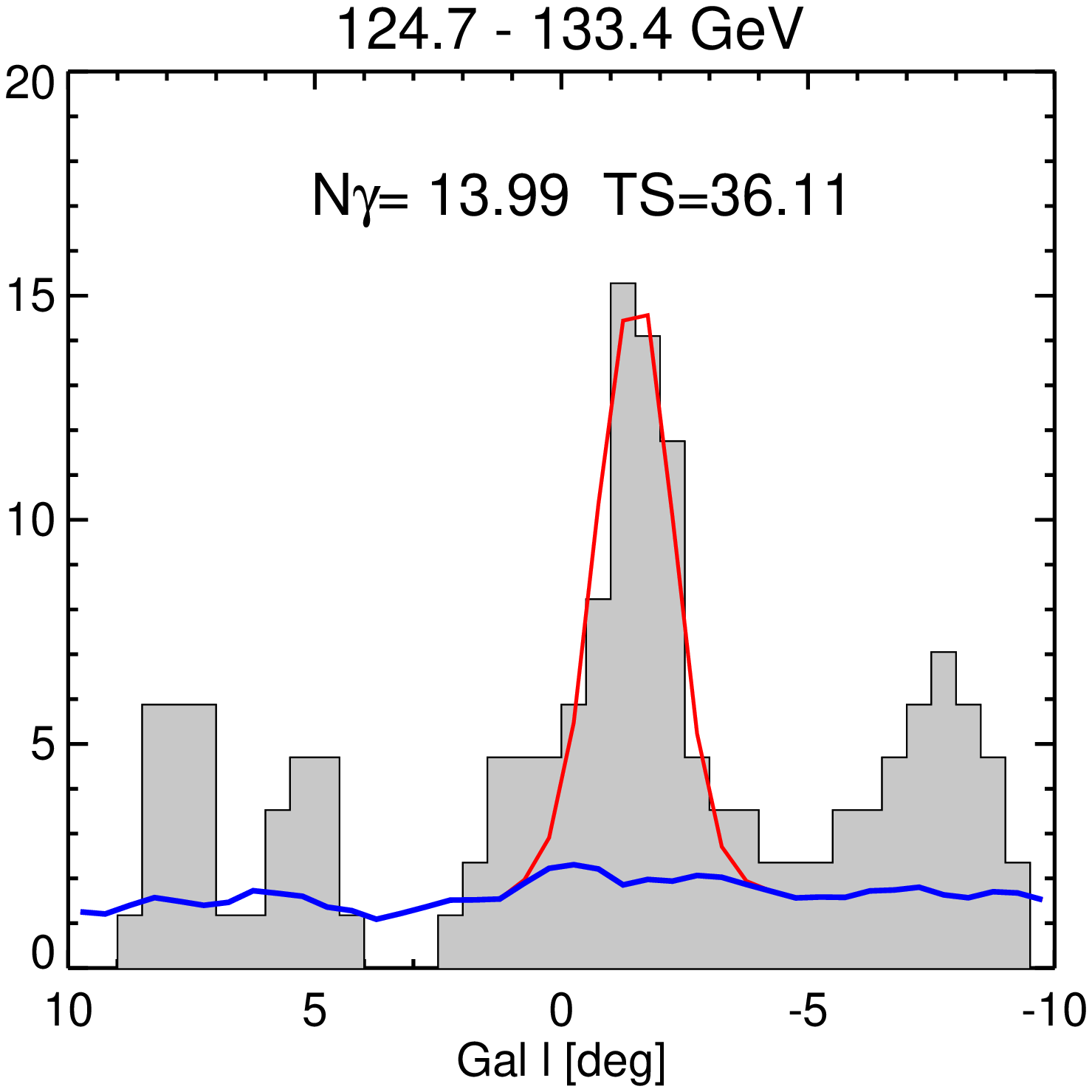}
        \includegraphics[width=0.36\textwidth]{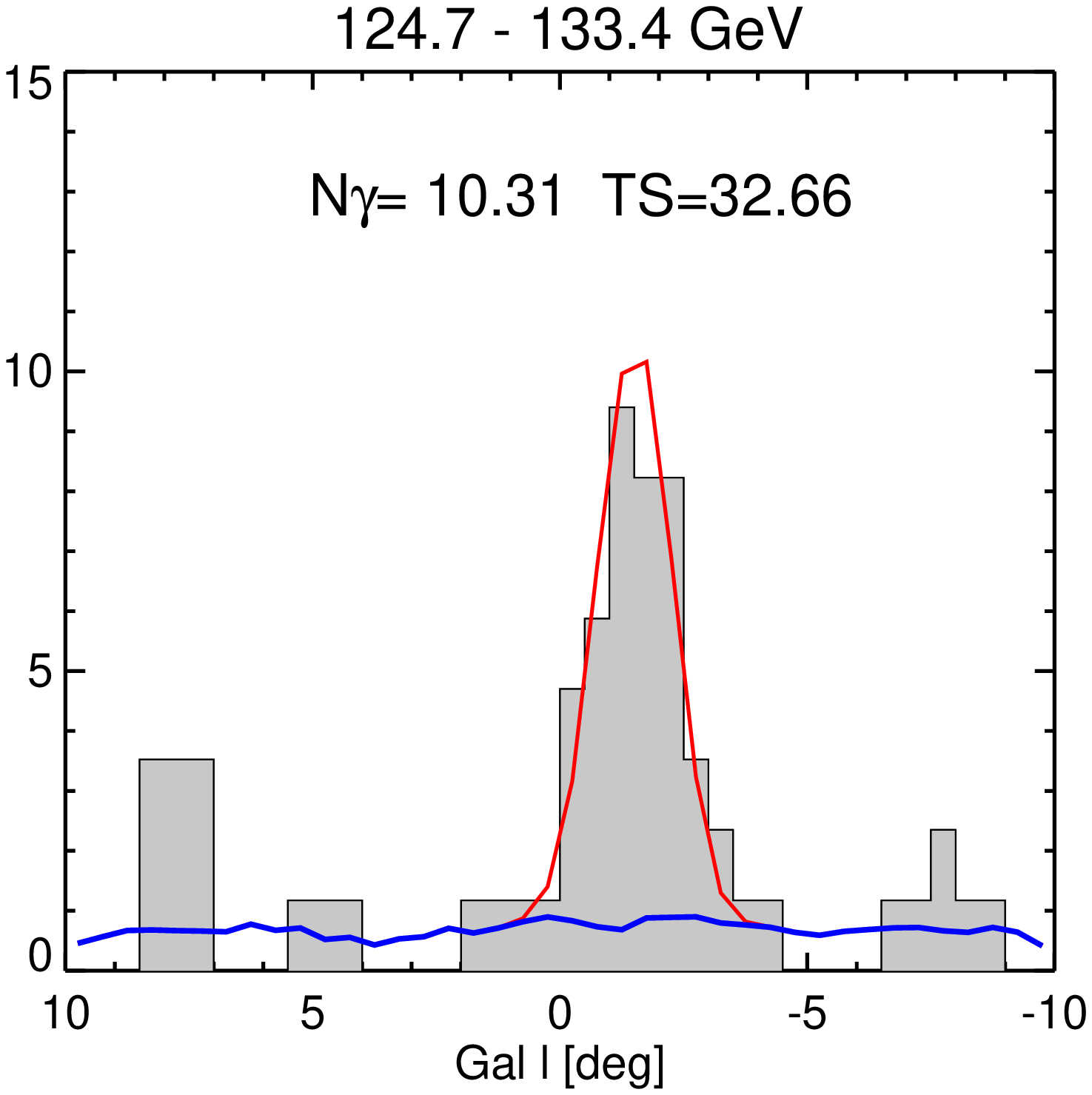}
        \includegraphics[width=0.36\textwidth]{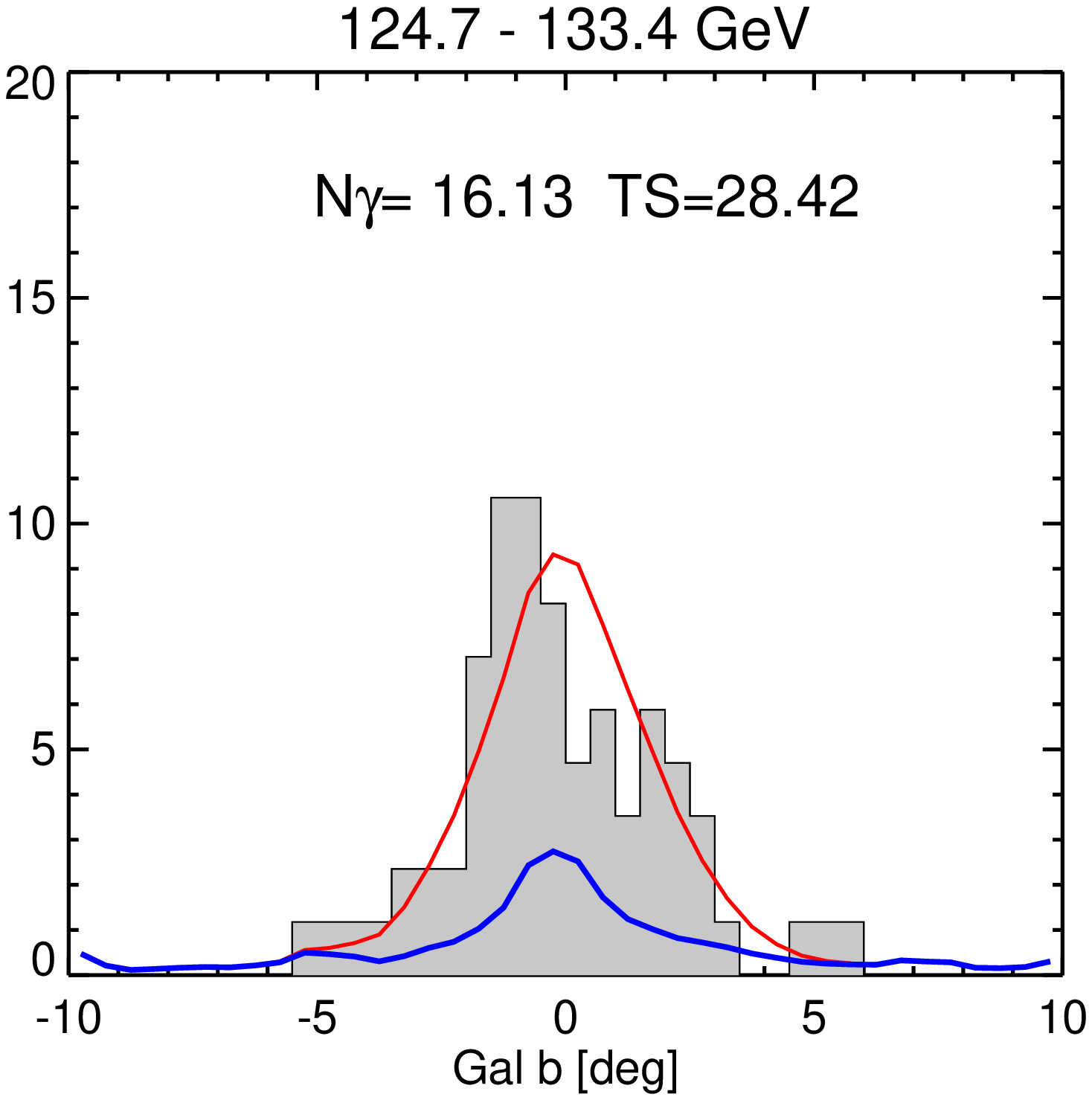}
        \includegraphics[width=0.36\textwidth]{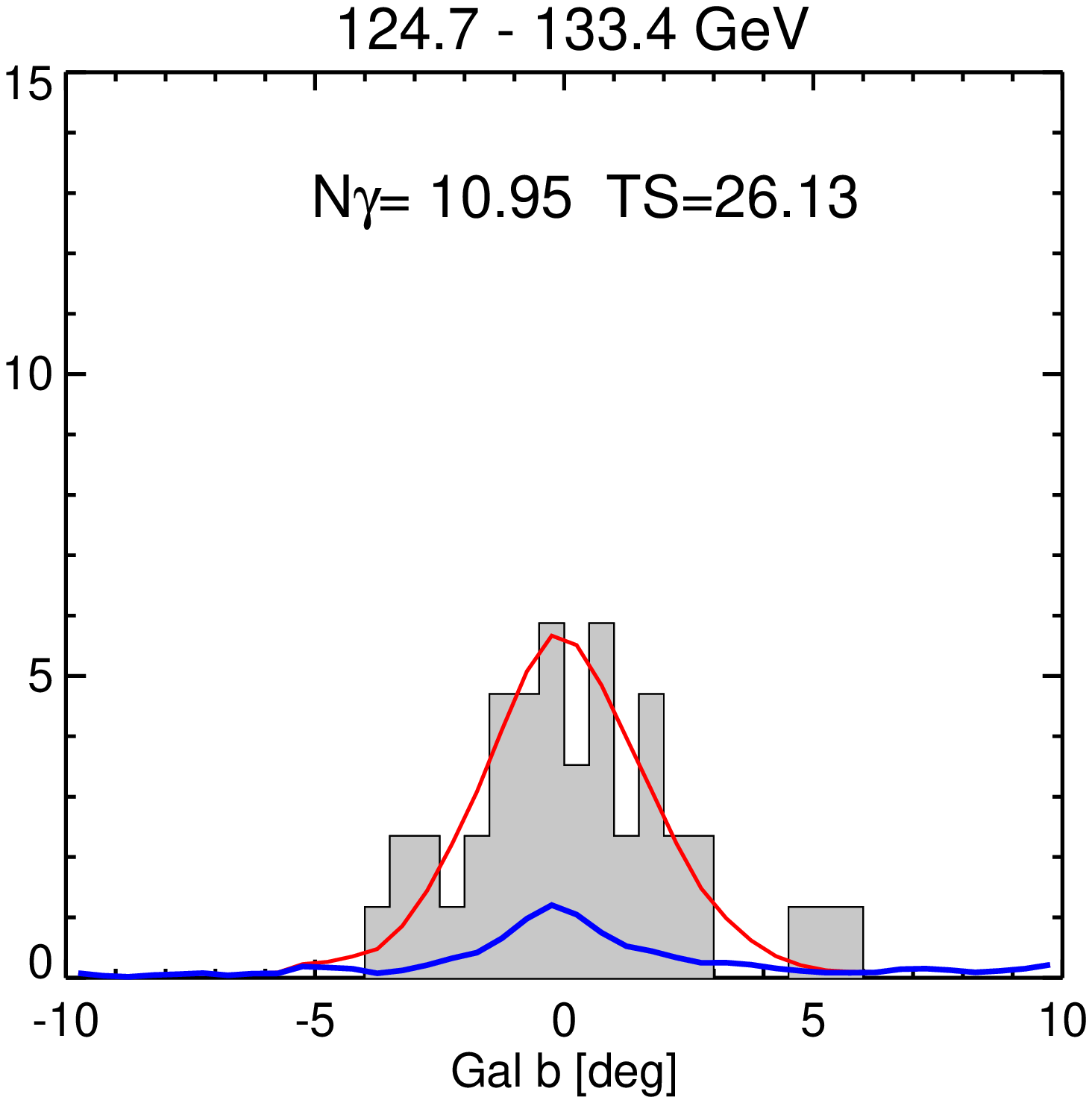}
    \end{center}
\caption{Profiles for both $\ell$ and $b$.  Even though the
  high-incidence-angle photons ($\theta > 40\degree$; right) panels have half
  the exposure (9.7\% vs. 19\% for the
  left panels), they have more than half of the photons, and nearly the same
  TS due to lower off-line background leaking in.  This demonstrates the
  statistical power of the high-incidence photons for line detection.  See
  section \ref{sec:profiles} for a discussion of the significance. 
}
\label{fig:4profiles}
\end{figure*}

\begin{figure*}[ht]
    \begin{center}
        \includegraphics[width=0.8\textwidth]{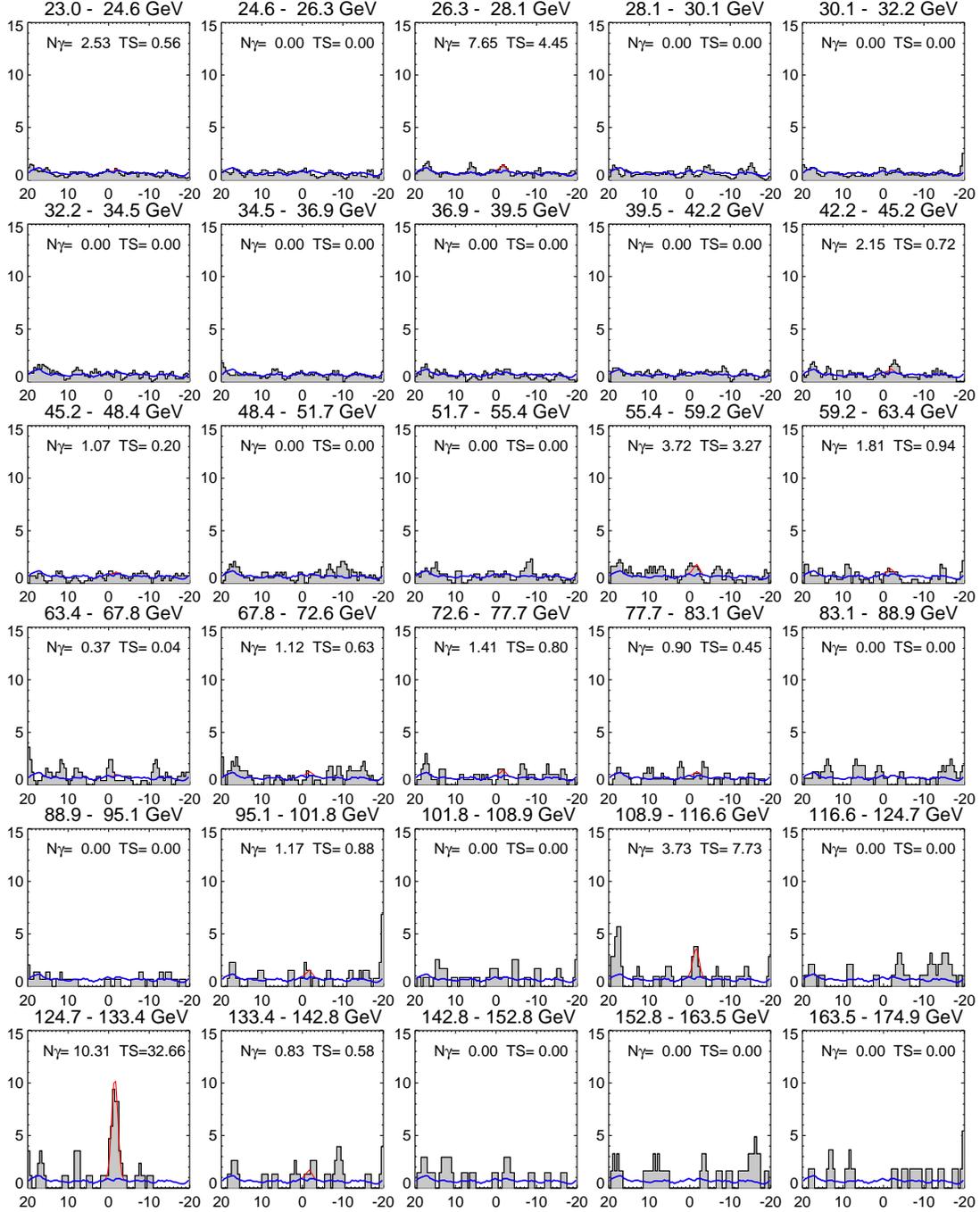}
    \end{center}
\caption{Profile of high-incidence photon ($\theta > 40\deg$) longitude
  distribution for $|b| < 5\degree$.  The $0.5\degree$ bins have been
  smoothed by a 3-bin box, and rescaled to arbitrary units of $E^{2.6} dN/dE$,
  making the background disk emission constant with $E$.  The background
  (\emph{blue}) is the average in these units for $10 < E < 50$ GeV.  In each
  panel, the (non-negative) amplitude of a FWHM$_\ell = 1.4\degree$ Gaussian
  centered at
  $\ell=-1.5$ is fit by maximizing the Poisson likelihood.  The corresponding
  number of photons and test statistic (TS) are displayed.  The only energy
  bin with significant emission is the 124.7-133.4 bin, centered on 129 GeV.
  See text for discussion of significance.  The bin centered on 113 GeV is not
  significant by itself, but is compatible with a line strength of 1/3 to 1/2
  that of the putative 129 GeV line. 
  }
\label{fig:30panels}
\end{figure*}

\begin{figure}[ht]
  \begin{center}
    \includegraphics[width=0.45\textwidth]{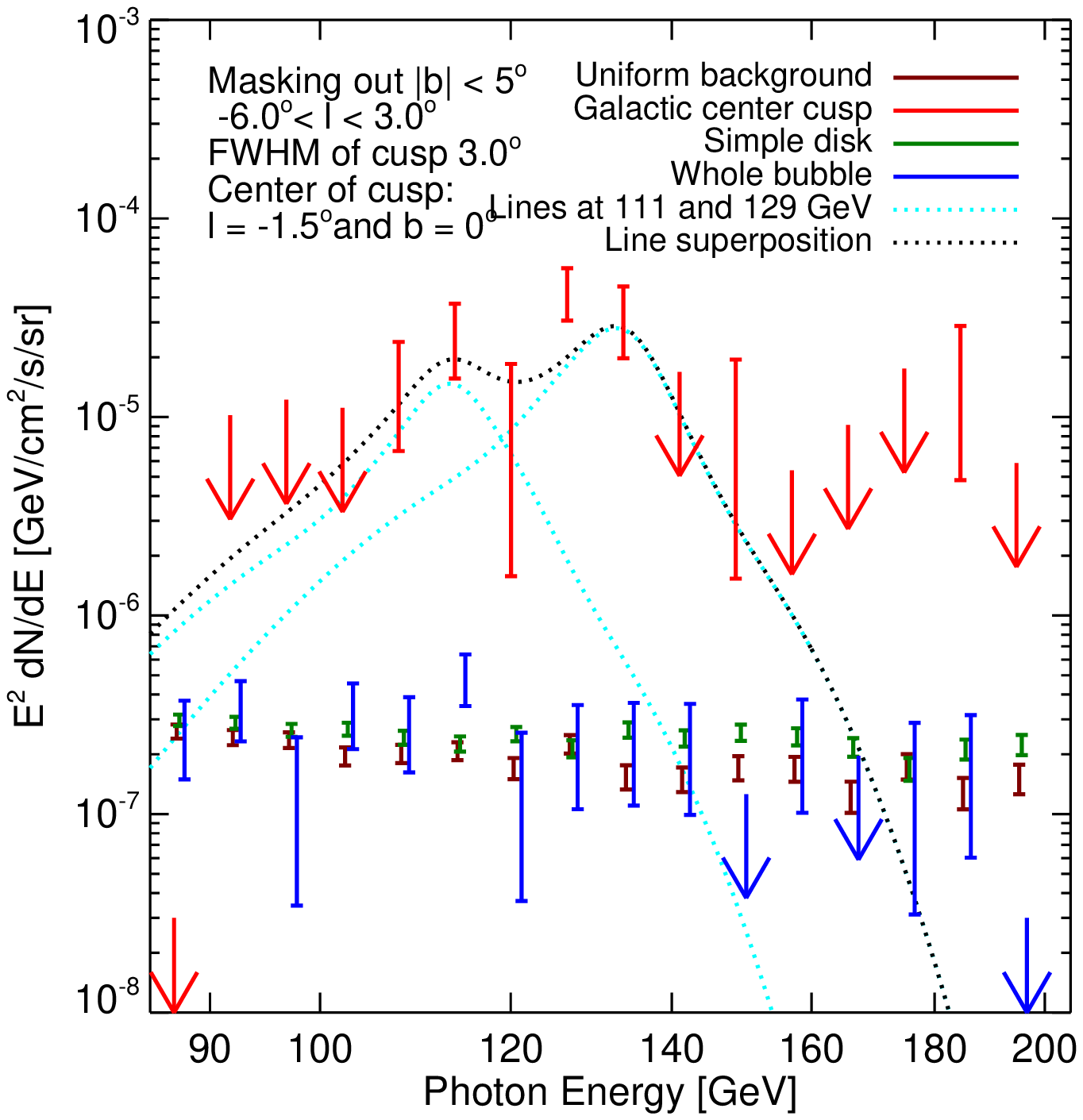}
  \end{center}
  \caption{Same as right panel of \reffig{fig9} but with
the cusp template centered at $\ell = -1.5\degree$ and
$b=0\degree$. 
  }
  \label{fig:fig16}
\end{figure}

\section{Detailed Spatial profile of the cusp}
\label{sec:profiles}

The cusp template used in \Refsec{cuspspec} is assumed to be centered on
the GC, and the choice of $4\degree$ FWHM is somewhat arbitrary.  We would
like the
data tell us what template to use, but there are so few photons, it is
difficult to make sense of an unsmoothed map of counts.  However, the smoothed
maps indicate the cusp is slightly off center (to the W of the GC) and it is
essential to follow this up. 

In this section, we consider individual photon events (not maps) and assume
the exposure across the GC is slowly varying.  We project the event locations
into histograms of $\ell$ and $b$ and study the distributions, finding
parameters of a best-fit Gaussian.  In this way, we may find the location,
shape, and significance of the 130 GeV feature in a way that is independent of
the previous sections, though somewhat less principled because we do not
explicitly use the exposure map.  We will not use the results of this section
to raise our claimed significance, but rather to emphasize that the cusp is
centrally concentrated, has a sharp spectrum, and is somewhat off center.

We model the background spectrum to be $dN/dE\propto E^{-2.6}$, with the
amplitude in each bin set by the 10-50 GeV average.  The exposure map is a
weak function of position and energy, and we neglect that variation in this
analysis.  An index of
$-2.5$ gives a significantly worse fit by overestimating counts at 100 to 200
GeV.  The $\pi^0$ emission is closer to $-2.7$ at lower energies, but $-2.6$
is a conservative choice, because assuming lower background at 130 GeV would
make any excess more significant.

We consider two spatial projections of the photon distribution in the inner
Galaxy.  For the longitude projection, we project the region $|\ell| <
10\degree$, $|b| < 5\degree$, in $0.5\degree$ bins of $\ell$ yielding a nearly
flat distribution (blue line in top panels of \reffig{4profiles}).  For the
latitude distribution, we project the region $-5 < \ell < 2$, $|b| < 10$, in
$0.5\degree$ bins of $b$, finding the emission near the plane dominates
(bottom panels of \reffig{4profiles}).  From this we immediately see that the
Galactic plane is much brighter than elsewhere, but the Galactic center is
\emph{not} particularly brighter than elsewhere in the plane at 10 to 50 GeV.

In order to test for the existence of a bump, we compare the 
$\ln {\mathcal L}_0$ for the null hypothesis to a model with an additional
Gaussian of FWHM $F_\ell$, centered at $\ell_0$ with peak height $A_\ell$.  We
compute $\Delta \ln {\mathcal L} \equiv\ln({\mathcal L}/{\mathcal L}_0)$ and
express results using the test statistic~\citep{Mattox:1996}, $TS=2\Delta\ln {\mathcal L}$.  The
test statistic plays the role $\Delta\chi^2$ would play in a Gaussian
problem.  

In comparing the $\ln {\mathcal L}$ of two models, one must account for the
fact that the two models reside in different parameter spaces.  In our case,
the null model space is a subspace of the other with 3 fewer parameters,
obtained by setting $A_\ell$ to zero.  In this case, the $TS$
distribution is simply the $\chi^2$ distribution for 3 degrees of freedom. 

Using photons from all incidence angles, the addition of a Gaussian
improves the $TS$ by 36.  This is \emph{not} a $6\sigma$ result because of the
3 additional degrees of freedom.  Rather, the probability that $TS$ would be
36 or higher is $p=7.5\times 10^{-8}$, corresponding to $5.25\sigma$ local
significance (not including the global trials factor).  The parameters of the
Gaussian are $F_\ell = 1.4^{+1.6}_{-0.4}, \ell_0=-1.5\pm0.3$, and an amplitude
corresponding to 14.0 photons (red line, \reffig{4profiles}). 

In the latitude direction, the fit is complicated by the concentration of
conventional continuum emission in the plane.  The cusp is not significantly
offset in the $b$ direction, but sits in the region of highest background, so
addition of the cusp is not demanded as strongly by the fit.  We introduce two
new degrees of freedom, the amplitude and FWHM of a Gaussian centered at
$b_0=0$.  This yields $TS=28.4$ and $p=6.8\times 10^{-7}$, corresponding to
$4.8\sigma$ (local significance).  The maximum likelihood parameters of the
Gaussian are $F_b=3.9^{+1.5}_{-0.7}$ and $A_b$ corresponding to 16.1 photons.
Both the $\ell$ and $b$ fits are roughly compatible with FWHM=$3\degree$, but
there is a slight preference for an elongation of the cusp in the $b$
direction.  A careful study of this will require much more data. 

In \reffig{4profiles} (right panels) we also display the same plots for the
high-incidence sample ($\theta > 40\degree$).  See \reffig{30panels} for such
plots in 30 energy bins.  The high-incidence-angle subsample contains half of
the exposure time (9.7\% vs. 19\%) but due to better energy resolution
($\Delta E/E\sim 0.06$) has less background on the line, and therefore yields
a $TS$ almost as large as the full data.  In this sense, \emph{most of the TS
  results from high $\theta$ events.}  This subsample would have yielded
$TS=32.6 (p=3.9\times10^{-7}, 4.93\sigma)$ for the $\ell$ profile, and
$TS=26.1 (p=2.2\times10^{-6}, 4.59\sigma)$ for the $b$ profile.  Although
these are slightly worse $p$ values than for the full data, they may actually
be more persuasive due to the lower background.

The fact that the cusp appears to be significantly off center implies that our
spectral fit in the previous section erred by using a centered cusp template. 
In \reffig{fig16} we show the measured energy spectrum of
a $3\degree$ FWHM cusp template, centered at $\ell=-1.5\degree$ and
$b=0\degree$. The local significance of this fit is $5.5\sigma$ relative to
the null hypothesis of zero intensity.  This improvement is heartening;
however, because of the extra parameter, the trials factor is now larger,
diluting the significance.

\begin{figure}[ht]
\begin{center}
\includegraphics[width=0.49\textwidth]{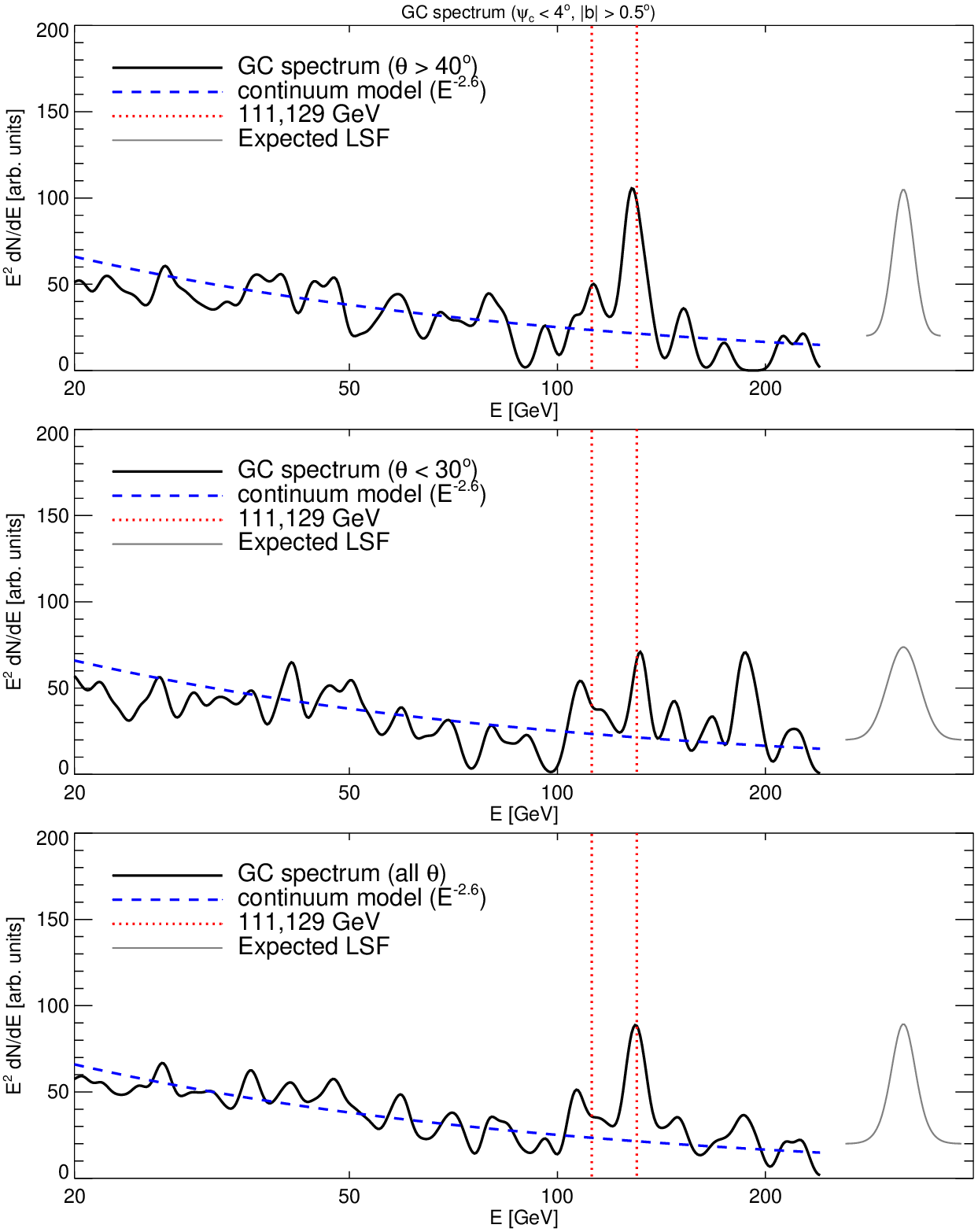}
\end{center}
\caption{
Spectrum of emission within $4\degree$ of the cusp center $(\ell,b)=(-1.5,0)$,
excluding
$|b| < 0.5\degree$.  High-incidence angle events (\emph{upper panel}) have a
factor of $\sim2$ better energy resolution than those that enter the
LAT close to normal incidence (\emph{middle panel}) or the whole sample
(\emph{lower panel}).  All three spectra have been
smoothed by a Gaussian of 0.06 FWHM in $\Delta E/E$, similar to the
expected resolution of the upper panel.  The continuum model is
$dN/dE\sim E^{-2.6}$, normalized at $20 < E < 50$ GeV (\emph{blue dashed}). 
}
\label{fig:unbinned}
\end{figure}

\begin{figure*}[ht]
    \begin{center}
	\includegraphics[width=0.45\textwidth]{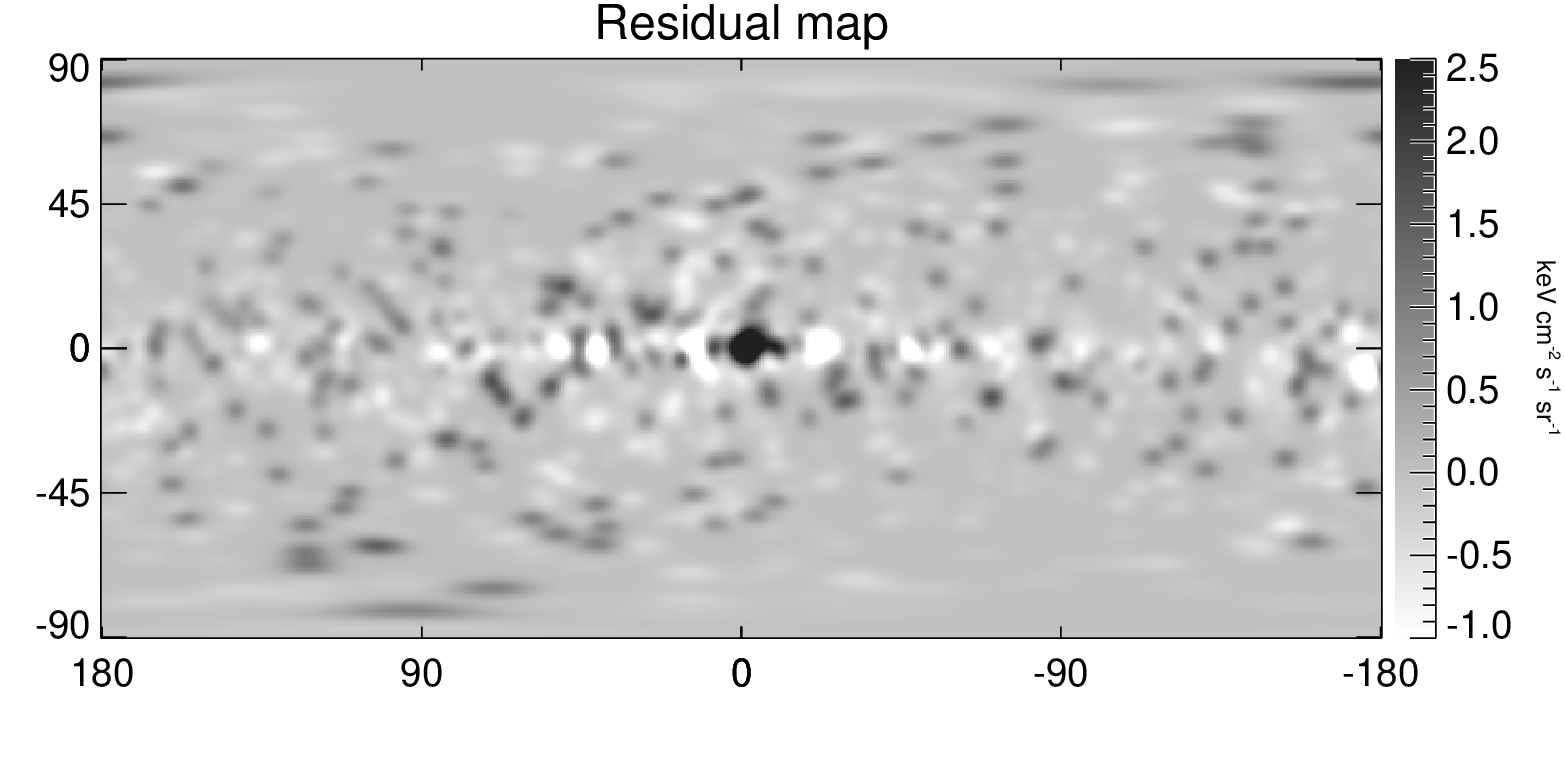}
	\includegraphics[width=0.45\textwidth]{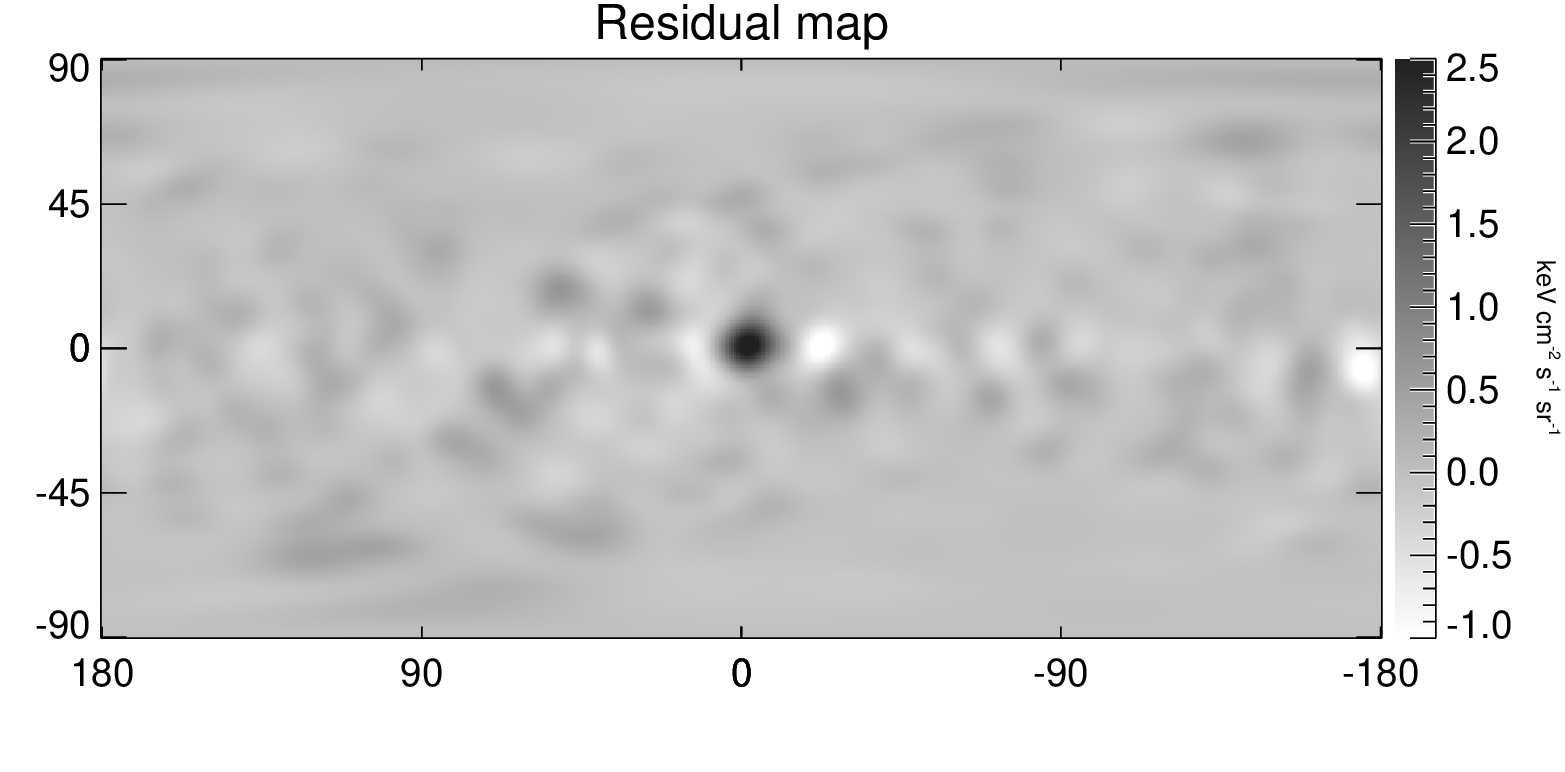}
	\includegraphics[width=0.45\textwidth]{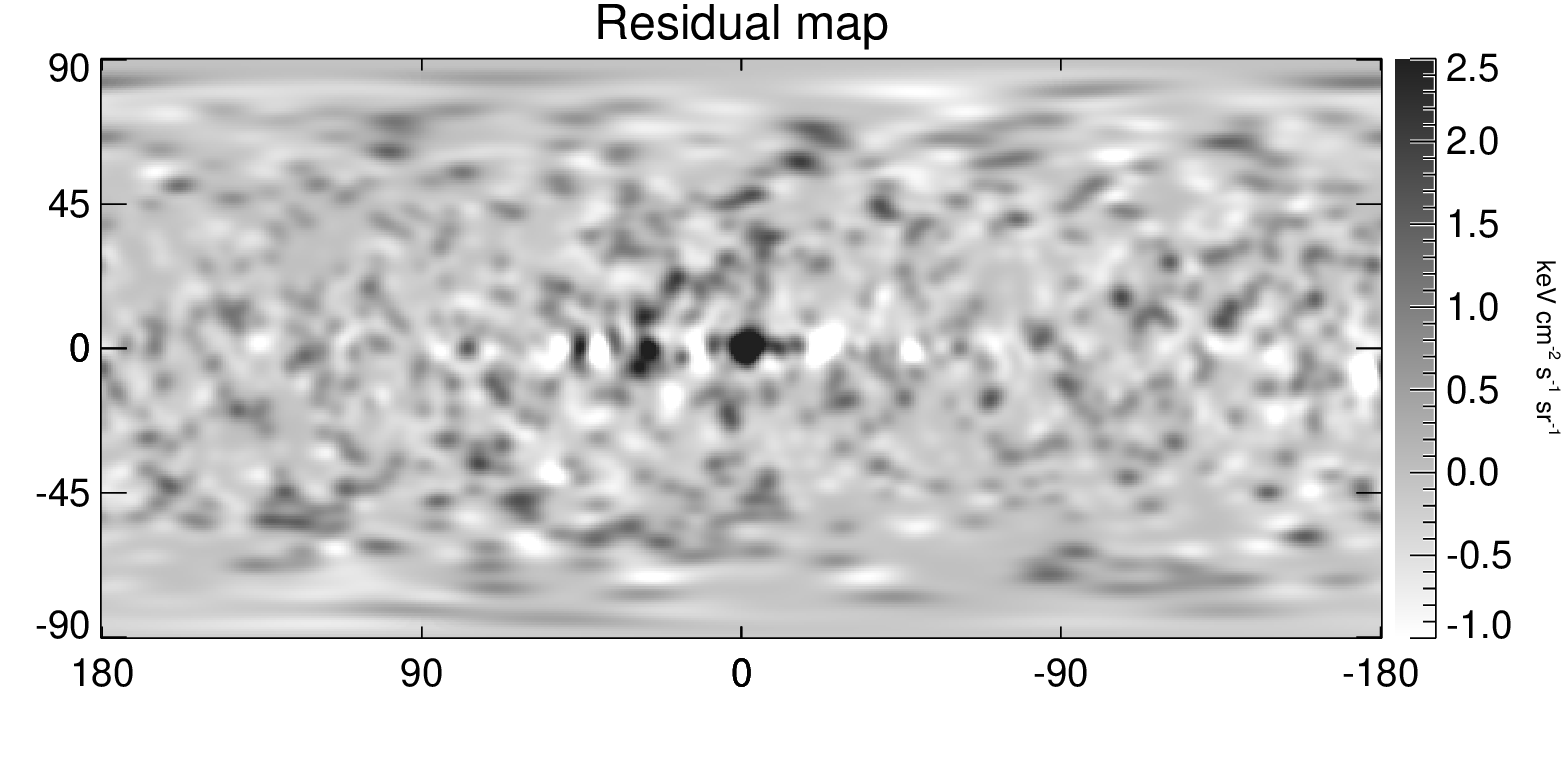}
	\includegraphics[width=0.45\textwidth]{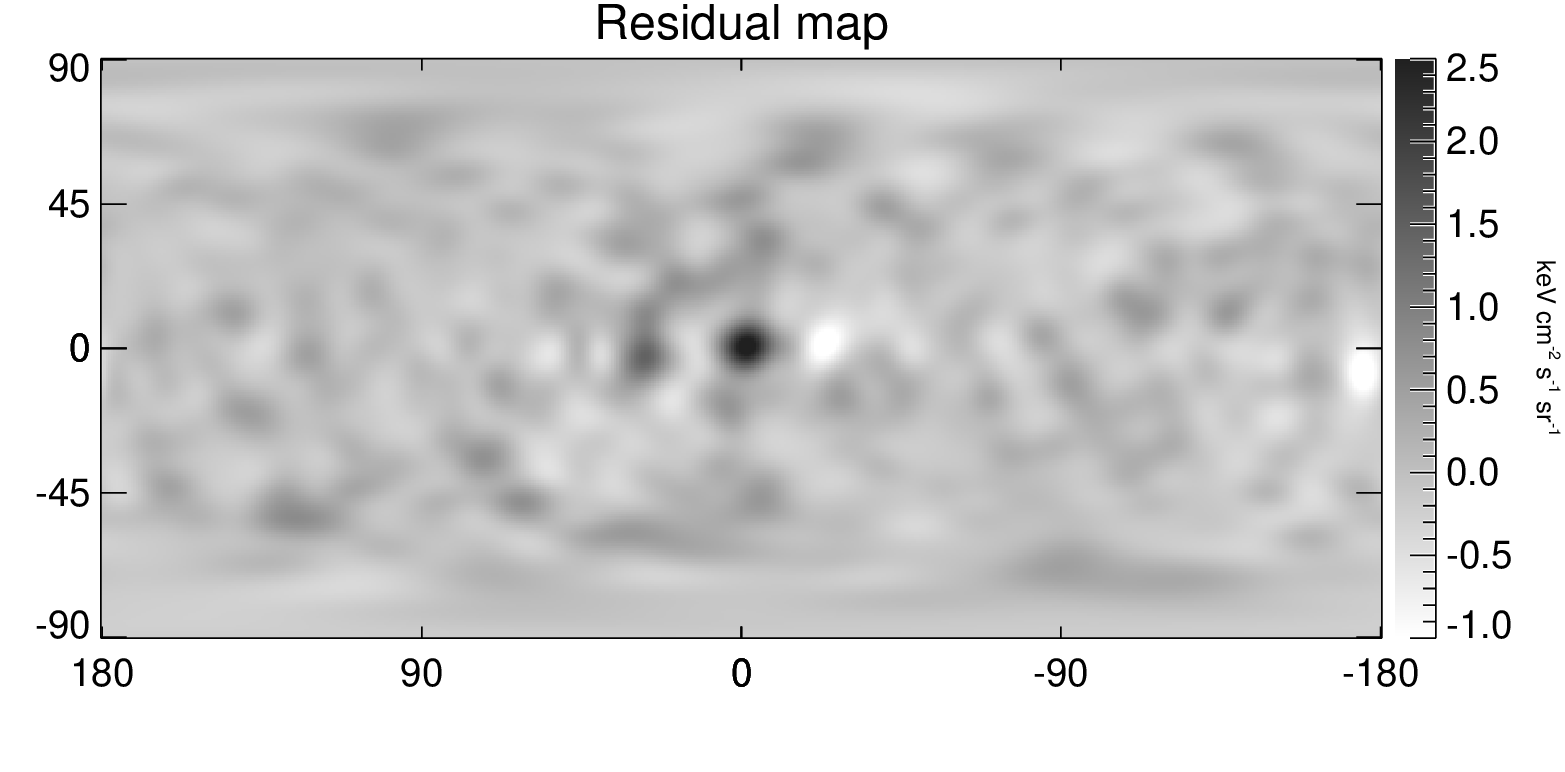}
    \end{center}
\caption{Same as \reffig{fig5}, but using only events with
high incidence angle $\theta > 40\degree$ which has better
energy resolution. The maps have been smoothed to 5$\degree$
(left two panels) and 10$\degree$ (right two panels), and
use \texttt{CLEAN} event class (upper two panels) and
\texttt{SOURCE} event class (lower two panels). The resolved
gamma-ray cusp structure is centered $1.5\degree$ W of the Galactic center. }
\label{fig:highincidence}
\end{figure*}

\begin{figure}[ht]
\begin{center}
\includegraphics[width=0.49\textwidth]{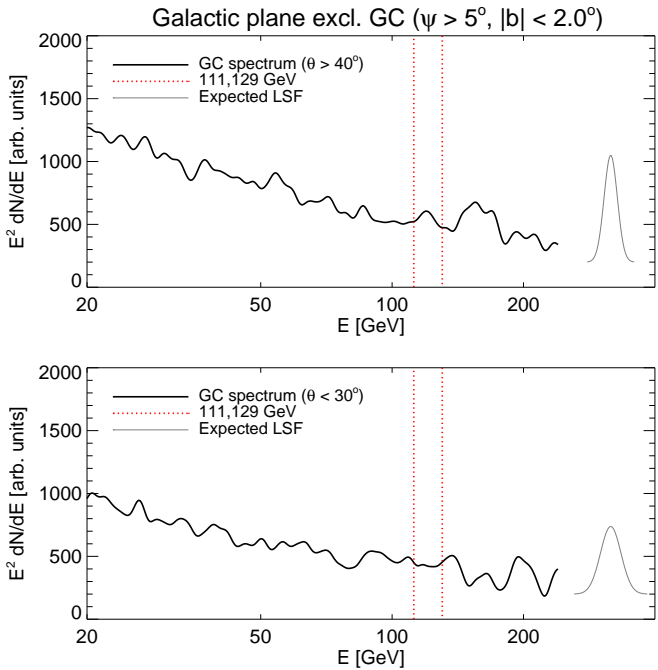}
\end{center}
\caption{
Like Figure \ref{fig:unbinned}, except for $|b| < 2\degree$ \emph{excluding}
photons within $5\degree$ of the Galactic center.  There is no indication of a
spectral feature near 130 GeV in the Galactic plane away from the Galactic
center. 
}
\label{fig:unbinnedplane}
\end{figure}

\begin{figure}[ht]
\begin{center}
\includegraphics[width=0.49\textwidth]{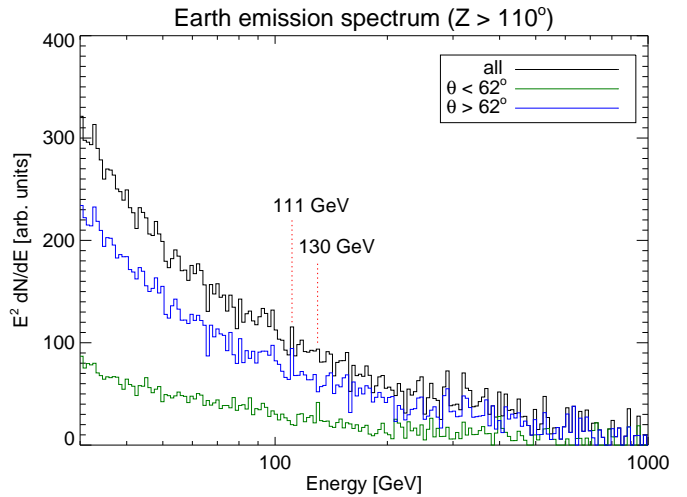}
\end{center}
\caption{
Photons from cascades in the Earth's atmosphere (sometimes 
incorrectly called ``albedo'' photons), and with $10\degree$ of the GC, show a
slight excess at 111 and 130 GeV also.  Because these photons arrive at
high zenith angle ($Z > 110\degree$), they tend to have a high incidence angle
(median $\theta=63.2\degree$).  The low-$\theta$ photons show a small bump at
130 GeV, and the high-$\theta$ photons show a small bump at 111 GeV.  The cuts
were chosen to maximize these features, so interpretation of this plot
requires a modest trials factor.
}
\label{fig:earth}
\end{figure}

\begin{figure*}[ht]
    \begin{center}
	\includegraphics[width=0.43\textwidth]{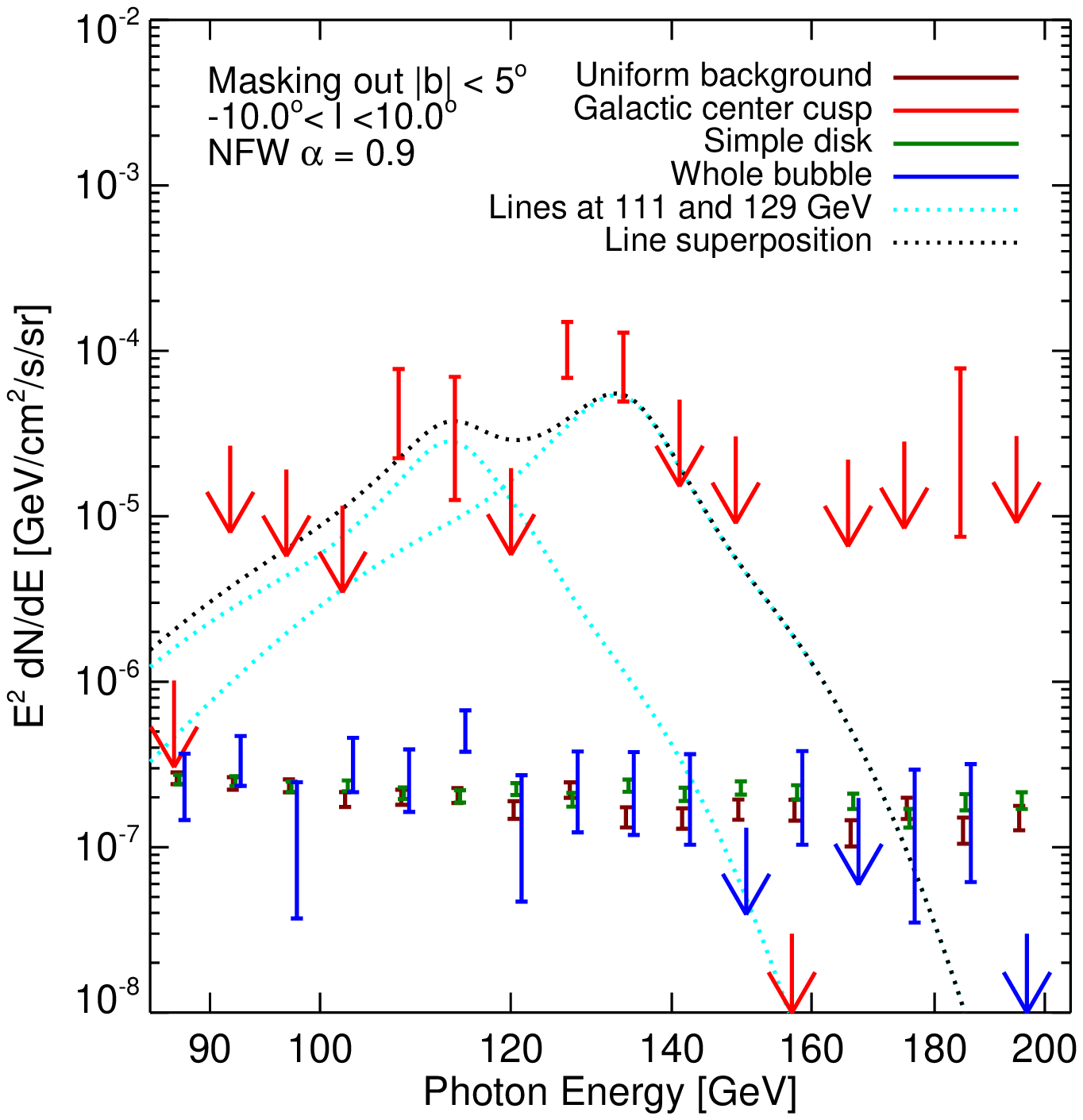}
	\includegraphics[width=0.43\textwidth]{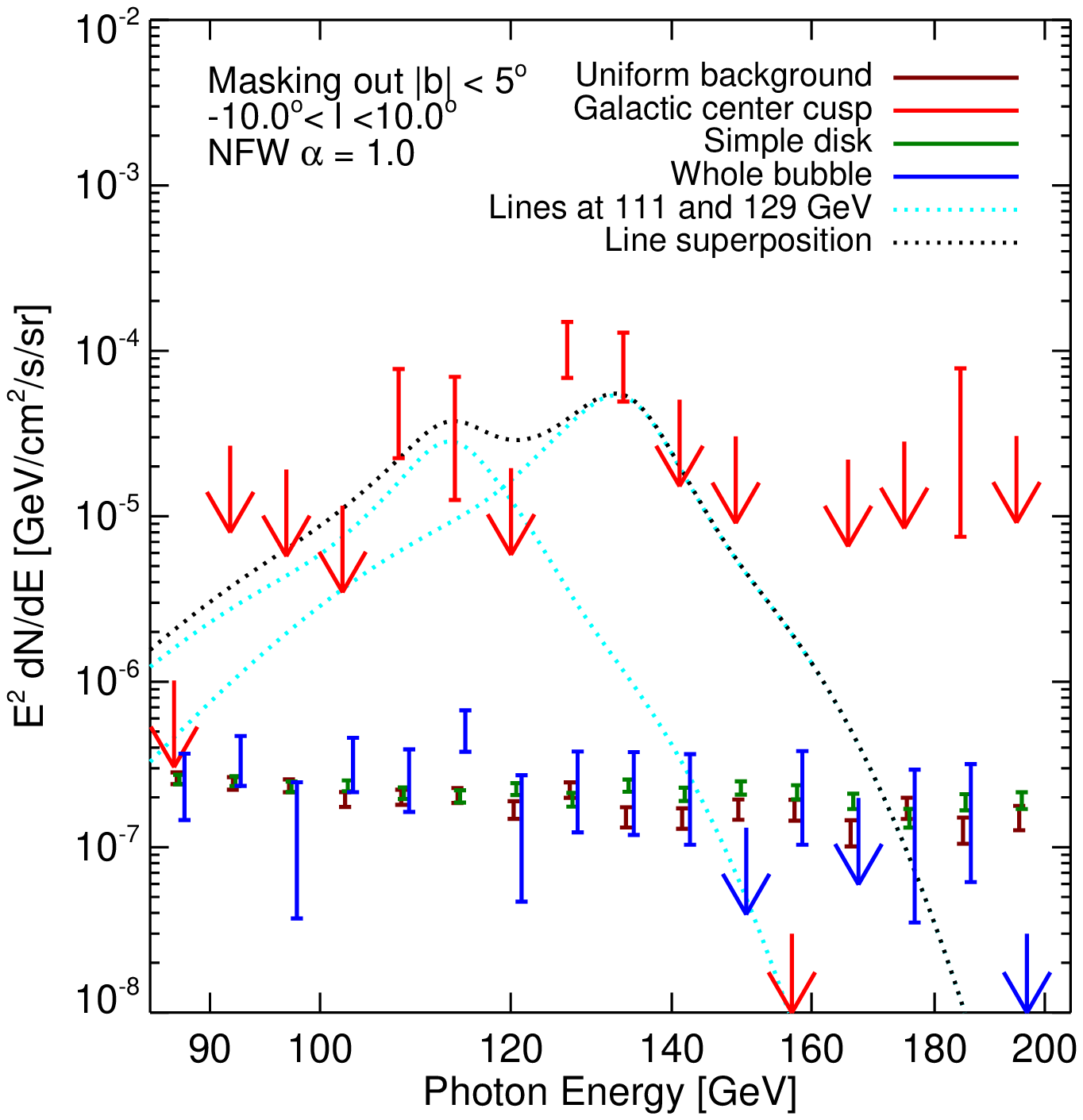}
	\includegraphics[width=0.43\textwidth]{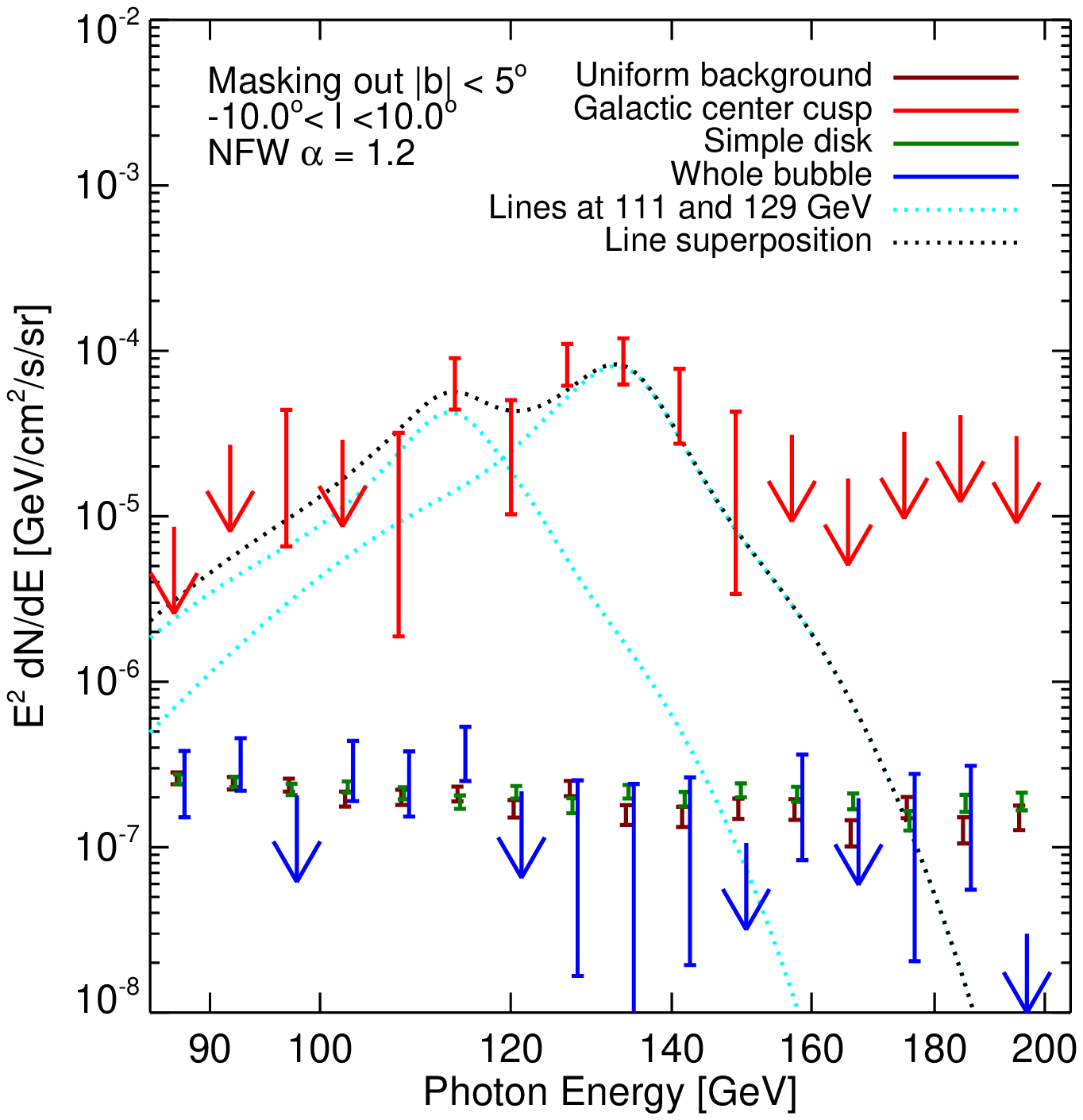}
	\includegraphics[width=0.43\textwidth]{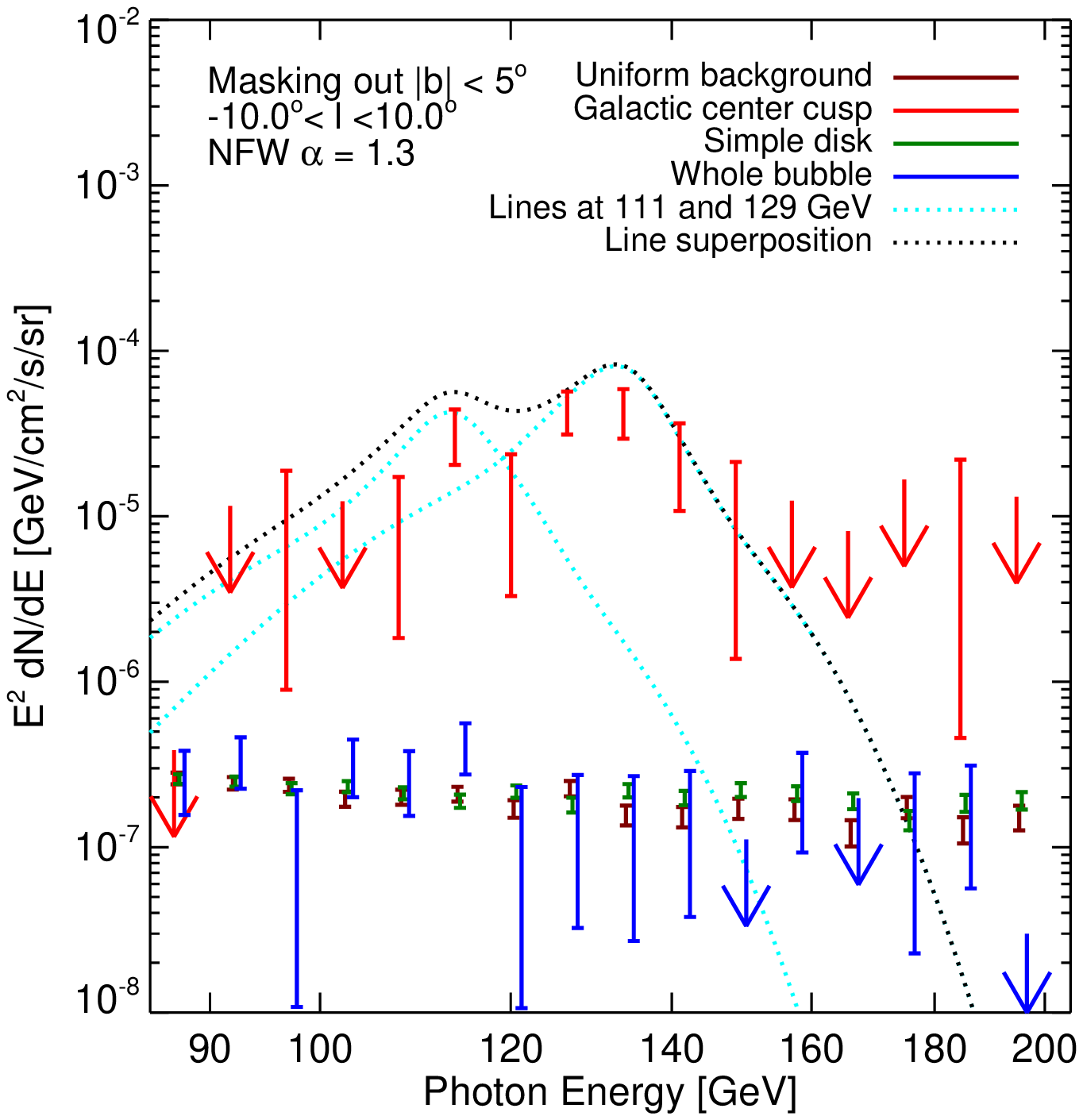}
    \end{center}
\caption{Same as right panel of \reffig{fig9} but using modified NFW profiles
  with inner power law index $\alpha$ for the DM density.  The density is
  squared and projected along the line of sight to produce the cusp template
  for, from
upper left to bottom right, $\alpha=0.9,1.0,1.2$, and $1.3$, 
respectively.  The fit prefers $\alpha=1.2$, yielding a
$6.5\sigma$ significance by combining the central 7 bins with
$>1\sigma$ each. }
\label{fig:fig20}
\end{figure*}

\begin{figure}[ht]
    \begin{center}
	\includegraphics[width=0.43\textwidth]{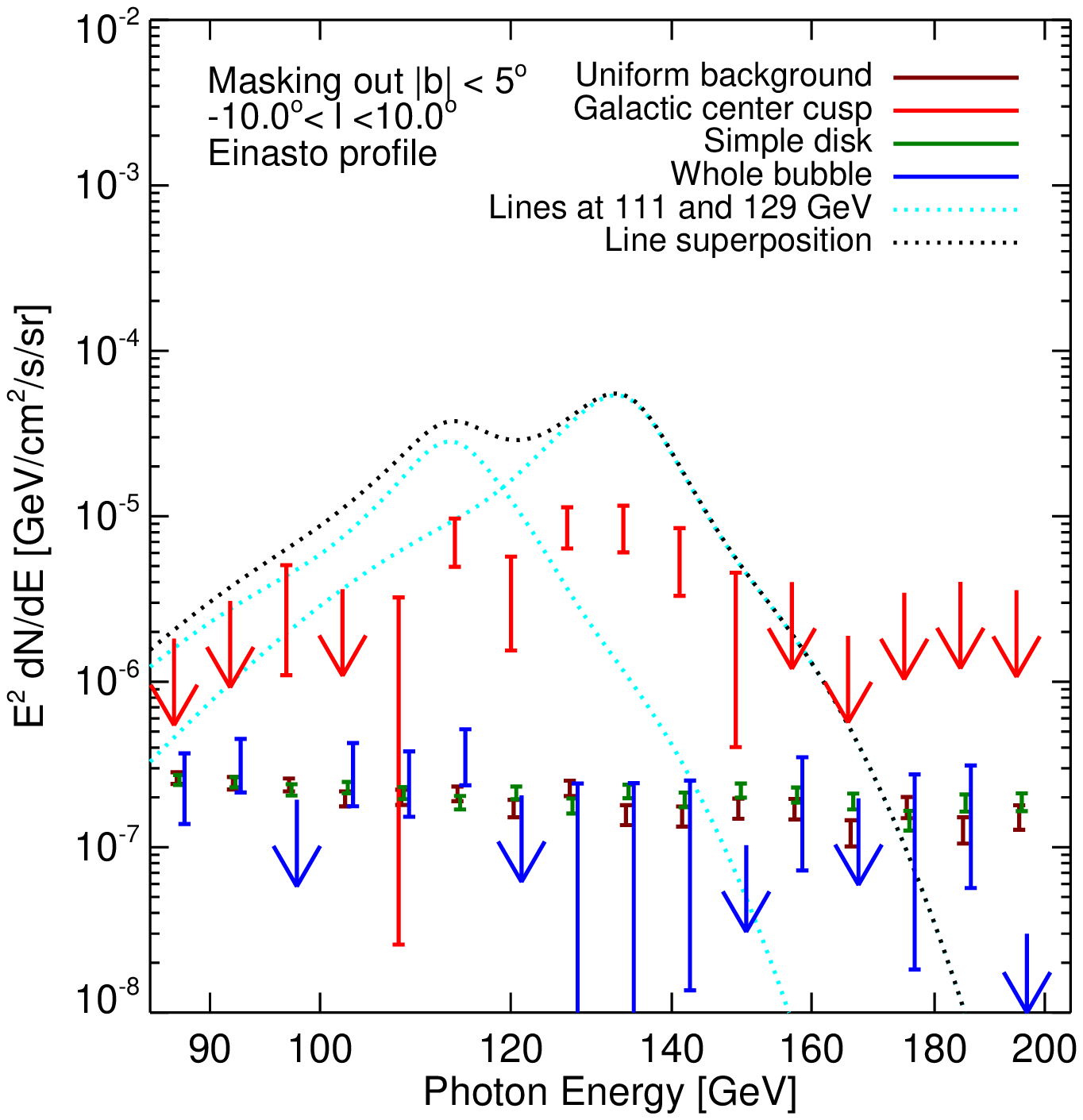}
    \end{center}
\caption{Same as right panel of \reffig{fig9} but using an
Einasto ($\alpha=0.17$) dark matter profile template to fit the
data, we obtain a $6.5\sigma$ detection of the cusp
only using the central 7 bins with $>1\sigma$ each. }
\label{fig:fig21}
\end{figure}

\section{Validation Tests}
\label{sec:validation}

\subsection{Assessment of line profile}
In section \ref{sec:cuspspec}, we investigated the cusp emission
by analyzing maps in various energy bins.  This allowed a separation of
spectral components by morphology, but relied on an arbitrary choice of
binning.  The result -- that there is a cusp of emission in the inner Galaxy
-- motivates an unbinned analysis of this region. 

In an unbinned analysis, one dispenses with arbitrary binning choices (size
and shift) and instead analyzes individual photon events.  For example, the
parameters
of a well defined model may be estimated with no binning in space or energy.
In the absence of a principled model, a compromise technique is to convolve
a finely binned
energy histogram with some kernel and compare profiles of prospective lines
with those expected for a true line, i.e. the instrumental response convolved
with the smoothing kernel.

In the case of LAT data this allows us to do an interesting reality check.
Energy
resolution of events at high incidence angle $(\theta\sim60\degree)$ is a
factor of $\sim2$ better than that of normal-incidence photons, motivating the
following test.

We select low incidence ($\theta<30\degree$) and high
incidence $(\theta>40\degree)$ photon samples.  We restrict to those
near the cusp center at $(\ell,b)=(-1.5,0)$ ($\psi_c < 4\degree$) but not in
the plane ($|b| < 0.5\degree$).  We then convolve each with a kernel and
compare them (Figure \ref{fig:unbinned}).  We adopt an LSF with a FWHM of
$\Delta E/E=0.06$ for high incidence and 0.12 for low
incidence \citep{Edmonds:thesis}, and in both cases convolve
with another FWHM 0.06 Gaussian.  After convolution, the LSF
is FWHM 0.085 for high incidence and 0.134 for low
incidence.  Normalized Gaussians of these widths are shown
for reference, normalized to the expected line strength at
130 GeV. Maps constructed using only high incidence events
are shown in \reffig{highincidence}.

Note that:
\begin{itemize}
\item The 129 GeV feature shape is strikingly similar to that expected for a
  line. The 111 GeV feature is unconvincing, but is also compatible with a line.
\item In some cases, fluctuations appear, but are not present in both low and
  high incidence spectra.
\end{itemize}
This analysis did introduce some additional parameters, but we have made
natural choices for them:  The 68\% containment radius of the cusp is
approximately $4\degree$, the Galactic ridge is about $0.5\degree$ thick, and
the $\Delta E/E = 0.06$ smoothing kernel is similar to the LSF of the LAT at
high incidence.  Smoothing a spectrum by its LSF is often a good compromise
between resolution and noise suppression in the high-noise limit.  Because
these parameters are all fixed to natural values, there is no significant
trials factor for this test, apart from the obvious one, that the lines could
have appeared anywhere (Section \ref{sec:trials}). 

This test did not have to succeed.  The fact that the high-incidence photon
sample has sharper spectral features is important; if the high-$\theta$ 
and low-$\theta$ spectra in
\reffig{unbinned} had been reversed, it would have been devastating for the
line hypothesis.

\subsection{Null test: Galactic plane spectrum}
To emphasize that the line feature in \reffig{unbinned} appears near the
Galactic center and not elsewhere, we perform the same analysis on the
Galactic plane ($|b| < 2\degree$) \emph{away} from the GC ($\psi >
5\degree$).  We find no indication of a line in either high-incidence or
low-incidence photons (\reffig{unbinnedplane}).

\subsection{Null test: Earth emission photons}
Another null test is provided by the Earth emission photons.  Cosmic-ray induced
cascades in the Earth's atmosphere shower photons on the LAT at high zenith
angle ($Z > 108\degree$).  These provide another null test, as there is no
reason for there to be a 130 GeV feature in the Earth emission spectrum.  On average,
no feature is seen (\reffig{earth}).  However, there is a hint of a line at
130 GeV in the low-incidence events and one at 111 GeV in the high-incidence
events.  These features also require a trials factor, but if they turn out to
be robust, then serious concerns would be raised about instrumental artifacts
giving rise to the observed lines.


 \begin{table*}
 \begin{center}
 \begin{tabular}{@{}rrrrrr}
\hline
\hline
Models & Before trials & After trials (one line) & Trials factor (one line)\\
\hline
Gaussian (centered)                        &    $5.0\sigma$ &    $3.7\sigma$ & 300  \\
Gaussian (off center, $\theta >40\degree$) &    $5.5\sigma$ &    $3.7\sigma$ & 6000 \\
unbinned $\ell$                            &    $5.2\sigma$ &    $3.2\sigma$ & 6000 \\
unbinned $\ell$ ($\theta >40\degree$)      &    $4.9\sigma$ &    $2.8\sigma$ & 6000 \\
unbinned $b$                               &    $4.8\sigma$ &    $3.5\sigma$ & 300  \\
unbinned $b$ ($\theta >40\degree$)         &    $4.6\sigma$ &    $3.2\sigma$ & 300  \\
NFW $\alpha=1.0$ (off center) &    $6.1\sigma$&    $4.5\sigma$& 6000  \\
NFW $\alpha=1.2$ (off center) &    $6.5\sigma$&    $5.0\sigma$& 6000  \\
NFW $\alpha=1.3$ (off center) &    $6.0\sigma$&    $4.4\sigma$& 6000  \\
NFW $\alpha=1.4$ (off center) &    $5.6\sigma$&    $3.8\sigma$& 6000  \\
NFW $\alpha=1.5$ (off center) &    $5.2\sigma$&    $3.2\sigma$& 6000  \\
{\bf Einasto (off center)}  & {\bf 6.6$\sigma$}& {\bf 5.1$\sigma$} & {\bf 6000} \\
\hline
 \end{tabular}
 \end{center}
\caption{The detection significance of the gamma-ray cusp
for various models.  See \refsec{trials} for a discussion of trials factors.}
\label{tbl:significancetable}
\end{table*}

 \begin{table*}
 \begin{center}
 \begin{tabular}{@{}rrrr}
\hline
\hline
Models & After trials (two line) & Trials factor (two line) \\
\hline
Gaussian (centered) &    $4.3\sigma$ & 36 \\
Gaussian (off center, $\theta >40\degree$) &   $4.2\sigma$ & 720 \\
NFW $\alpha=1.0$ (off center) &   $4.9\sigma$& 720 \\
NFW $\alpha=1.2$ (off center) &   $5.4\sigma$& 720  \\
NFW $\alpha=1.3$ (off center) &   $4.8\sigma$& 720  \\
NFW $\alpha=1.4$ (off center) &   $4.3\sigma$& 720  \\
NFW $\alpha=1.5$ (off center) &   $3.8\sigma$& 720  \\
{\bf Einasto (off center)}          & {\bf 5.5$\sigma$}  & {\bf 720} \\
\hline
 \end{tabular}
 \end{center}
\caption{The detection significance of the gamma-ray cusp
structure with different models.  See \refsec{trials} for a discussion of trials factors.
Masking out $0.5\degree$ around the GC area does not affect the
results.
}
\label{tbl:significancetable2}
\end{table*}

\section{fitting with Dark matter profile}
\label{sec:nfw}

In previous sections, we have modeled the line emission cusp with a Gaussian
in an attempt to keep the analysis generic.  However, WIMP annihilation is the
most likely explanation for gamma-ray line emission, and we would like to know
if the signal expected from a reasonable DM profile fits the data better than
a Gaussian.  In this section, we repeat our template analysis with an Einasto
profile as well as a set of modified NFW profiles.  We also estimate the dark
matter annihilation rate $\sigmav$ required for each model. 

The gamma-ray intensity in a given direction is the line-of sight integral of
the dark matter number density, squared,  along a given direction,
\begin{equation}\label{eq:flux}
\frac{d\Phi_\gamma}{dE_\gamma} = \frac{1}{8\pi} \, \frac{\langle \sigma v\rangle_{\gamma\gamma}}{m_\chi^2}\,
2\delta (E- E_\gamma)\, r_\odot \rho_\odot^2\, J,
\end{equation}
with: 
\begin{equation}\label{eq:J}
J = \int_{\rm cusp} db \int_{\rm cusp} d\ell \int_{\rm l.o.s.} \frac{ds}{R_\odot} \, \cos b\, \Bigg(\frac{\rho(r)}{\rho_\odot}\Bigg)^2,
\end{equation}
where the integral is over the cusp structure,
$\langle\sigma v\rangle_{\gamma\gamma}$ is the partial
annihilation cross-section for $\chi\chi \rightarrow
\gamma\gamma$, $m_\chi$ is the WIMP mass
($E_\gamma=m_\chi$), $R_\odot \simeq 8.5$ kpc is the distance
from the Sun to the GC \citep{Ghez:2008}, $\rho(r)$ is the
WIMP halo profile, $\rho_\odot \simeq 0.3$ GeV cm$^{-3}$ is
the often-used WIMP density at the Solar system \citep{Jungman:1995df},
$r = (s^2 + R_{\odot}^2 - 2 s R_\odot \cos\ell \cos
b)^{1/2}$ is the Galactocentric distance and $s$ is the line
of sight distance.  We use $\rho_\odot \simeq 0.3$ GeV cm$^{-3}$ to facilitate
comparison with earlier work, but a higher value $\rho_\odot \simeq 0.4$ GeV
cm$^{-3}$ \citep{Catena:2009mf} would reduce our $\sigmav$ values by a factor
of 1.7. 

For halo profiles $\rho(r)$, we consider the
Navarro-Frenk-White (NFW) profile~\citep{NFW},
\begin{equation}
\rho_{\rm NFW}(r) = \frac{\rho_s}{(r/r_s)^{\alpha}(1+r/r_s)^{3-\alpha}}
\end{equation}
with $r_s=20$ kpc. The value of $\rho_s$ is determined by
requiring $\rho(r_\odot) = 0.3$ GeV cm$^{-3}$. In
\reffig{fig20}, we show the energy spectrum of the
cusp structure using the NFW profile with various values of $\alpha$.
Interestingly, we obtain a $6.5\sigma$ ($5.6/5.9\sigma$ for
one/two line case after trials factor correction) detection
of the cusp assuming $\alpha=1.2$.  This corresponds to a $p$ value $10^4$
times smaller than $5\sigma$. 

We also consider the case of the Einasto profile,
\begin{equation}
\rho_{\rm Einasto}(r) = \rho_s \exp\{ -(2/{\alpha_E})[(r/r_s)^{\alpha_E} - 1]\}
\end{equation}
with $r_s=20$ kpc and
$\alpha=0.17$~\citep{Einasto:1965,Navarro:2004}. This
template provides a slightly better fit, with $6.6\sigma$
($5.7/6.0\sigma$ for one/two line case after trials factor
correction) detection of the cusp.
\reftbl{significancetable} and \reftbl{significancetable2}
lists the resulting significance for all the models.

\section{How to confirm this signal quickly: A Modified survey strategy}
\label{sec:strategy}
The fact that high-incidence-angle photons are superior for line detection
raises an exciting possibility: the scan
strategy of \Fermi-LAT could be altered for 1 year to confirm the 130 GeV line
(if real) at
$5\sigma$ with no trials factor, a significance that would be widely regarded
as a discovery.  If more than one line is present, the additional data would
help characterize it. 

\Fermi\ has usually scanned the sky in survey mode, observing the full sky
every 2 orbits with occasional slews to targets of opportunity.  This strategy
is excellent for uniformity of full-sky coverage, but is far from optimal for
collecting high-incidence-angle photons from the GC. 

From the spacecraft data files for the first 3.7 years (through week 202), it
is a straightforward exercise to derive the fraction of the time the GC is
accessible.  We impose the constraints that the roll angle be within
$35\degree$ of zenith (as in standard survey mode) and that the GC have an
incidence angle of $45\degree < \theta < 55\degree$.  We consider only times
when the spacecraft is not in the South Atlantic Anomaly (SAA) and the
\texttt{DATA\_QUAL} flag is good.  These constraints allow the GC to be
observed 40.6\% of the time.  This exceeds the exposure time of our
($40\degree < \theta < 60\degree$) sample (observed 9.7\% of the time) by more
than a factor of 4. In other words, LAT could gather high $\theta$ photons
from the GC $4\times$ faster than it usually does with a simple change to the
observing strategy.

Because we wish to point $\sim50\degree$ away from the GC anyway, it would
still be possible to maintain coverage of the inner $110\degree$ radius about
the GC.  On the other half of the sky, survey mode would continue as usual. 
Although this change may be sub-optimal for some science projects, it would be
a net benefit to the entire inner Galaxy, home to a great many scientific
objectives. 

We believe the trade would be worth it.  After 1 year of altered observing, we
would have a sample of high incidence photons equal to the current sample, and
could evaluate their significance directly, in the absence of any trials
factor.  In addition to confirming the line at 130 GeV, other lines (113 GeV)
may also become significant with additional data.  At this point in the LAT
mission, there are few ways for the instrument to dramatically improve its
sensitivity to new physics in its remaining lifetime.  This is a golden
opportunity.

\section{Discussion and Conclusion}
\label{sec:conclusion}

\emph{Morphology:}
Using 3.7 years of \Fermi-LAT data, we have
identified a resolved gamma-ray cusp structure toward the
Galactic center region.  To reveal this structure in a simple,
model-independent way, we first take a linear
combination of smoothed maps that cancels out continuum
emission in the plane and inspect the residual. We find that this structure only
appears in the energy range from $\sim 120$ GeV to $\sim 140$
GeV after searching $80 < E < 200$ GeV maps  (Figures
\ref{fig:fig3}-\ref{fig:fig5}, and \ref{fig:highincidence}).
The FWHM of the cusp morphology, if modeled with a Gaussian, is $\lesssim
4\degree$ and is unrelated to the
\Fermi\ bubble structure~\citep[as suggested by
][]{Profumo:2012}. No other region of the sky
reveals any significant excess in this energy range.


\emph{Template regression:} We perform a Poisson likelihood
analysis to obtain the energy spectrum of the gamma-ray
cusp.  We create maps in energy bins 5.5\% wide, and model
each map as a linear combination of templates including the
cusp template, a uniform gamma-ray background, a thin
gamma-ray disk tracing the Galactic plane, and the \Fermi\
bubble structure. By modeling the data with a linear
combination of templates and maximizing the Poisson
likelihood of observing the observed counts, we are able to
separate the cusp emission from the \Fermi\ bubbles and the
Galactic disk.  To the extent that the template is actually
correct, this ``matched filter'' gives an optimal estimate
of the flux in each energy bin.  The uncertainty estimates
include marginalizing over the uncertainty in the other
templates coefficients. We do not make \emph{a priori}
assumptions about the dark matter profile.  For a template
centered on the GC, we find that the cusp emits gamma rays
with a luminosity of $(3.2\pm0.6)\times 10^{35}$ erg/s, or
$(1.7\pm0.4)\times 10^{36}$ photons/sec. The null hypothesis
of zero intensity is ruled out by $5.0\sigma$ ($3.7\sigma$
with trials factor).

Motivated by the apparent offset between the cusp and the GC, we repeat the
template fit with a cusp template centered at ($\ell,b$) = (-1.5,0)
(\reffig{fig16}), with the local significance rising to $5.5\sigma$, but with
a larger trials factor, diluting the global significance.

\emph{Line profile:} The energy spectrum of the cusp
structure at $E \gtrsim$ 80 GeV is consistent with a single
spectral line (at energy $127.0\pm 2.0$ GeV with
$\chi^2=4.48$ for 4 d.o.f.) convolved by the LAT energy
response \citep{Rajaraman:2012}.  A pair of lines at
$110.8\pm 4.4$ GeV and $128.8\pm 2.7$ GeV provides a
marginally better fit (with $\chi^2=1.25$ for 2 d.o.f.). 
These line energies suggest a WIMP annihilating to $\gamma \gamma$ and
$\gamma Z$.  Fitting the two lines jointly, we find a WIMP mass of $127.3
\pm 2.7$ GeV with $\chi^2=1.67$ for 3 degrees of freedom. 
The line pair is also compatible with a 141 GeV WIMP annihilating
through $\gamma Z$ and $\gamma h$ for $m_h\sim125$ GeV, as
in the ``Higgs in Space''
scenario~\citep{Jackson:2010}. We note that the uncertainty of overall absolution energy scale of LAT is [-10, +5]$\%$~\citep{FermiInstrument}.

\emph{Interpretation as dark matter: } Given the properties
of the gamma-ray cusp, and assuming it originates from dark
matter annihilation, constraints on the dark matter density
profile and annihilation rate or cross section can be placed
on various models~\citep[e.g.][]{Goodman:2011, Jackson:2010,
Cline:2012, Bertone:2009, Dudas:2012,Geringer-Sameth:2012}.
At present, the extrapolation from the local dark matter
density into the Galactic center is uncertain enough that
there are large uncertainties in the cross section.  We do
not attempt to improve those estimates in this work.

\emph{Comparison to previous work:} The recent study by
\cite{2012arXiv1205.6474T} put constraints on line emission
from the inner Galaxy, excluding photons within $|b| <
5\degree$.  Therefore, their constraints are not in conflict
with the signal claimed in this paper, which is mostly
within $5\degree$ of the GC.  Ackermann et al. searched for
dark matter gamma-ray line signal from 4.8 GeV to 264
GeV~\citep{2012arXiv1205.2739F}, updating the results from
the earlier study with 11 months \Fermi-LAT
data~\citep{PhysRevLett.104.091302}. The results from this
study are obtained from two years of \Fermi-LAT data and
used the Pass 6 processing, which has somewhat worse
background and instrumental systematics compared to Pass 7.

The Galactic center is known to be a region with diffuse high
energy gamma-ray emission~\citep{HESS:2006}, especially the
Galactic center ridge~\citep{HESSGCRidge}. High energy gamma
rays are produced by cosmic rays interacting with
interstellar gas or giant molecular clouds in the
central $\sim$200 parsecs. We note that the resolved cusp is
incompatible with a point source given that it extends to
$\sim 4\degree$. To check for contribution from
unresolved point sources, we have repeated our analyses
masking out regions with $|b| < 1\degree$, and have
obtained similar results, confirming that the cusp is not 
associated with the Galactic ridge. 



\emph{The future: } The next version of \Fermi-LAT data
(Pass 8) will move us closer to realizing the full
scientific potential of the LAT~\citep{Atwood:2012}. The
expected improvements include reduced backgrounds, increased
effective area, reduced point-spread function, better
understanding of the systematic uncertainties, and
particularly extending the energy reach to higher photon
energy. Finally, the calorimeters of LAT have been
demonstrated to effectively operate as a standalone
instrument providing imaging of the sky at $E \gtrsim 10$
GeV, albeit with worse image resolution (at the level of $1
\degree$) with a large potential increase in the effective
area at high energy and large angle~\citep{Atwood:2012}.
Such data would be valuable for future studies of the gamma
ray line(s).

\emph{A modified survey strategy:} As discussed in section
\ref{sec:strategy}, the LAT could acquire significance on
spectral line emission from the Galactic center 4 times as
fast as it currently does with a change in observing
strategy.  In other words, a single year of observing the GC
40\% of the time would yield more significance on these
gamma-ray lines than all data currently in hand.  A factor
of 2 improvement would come from observing the GC more
often, and another factor of 2 by maintaining a high
incidence angle ($45\degree < \theta < 55\degree$), yielding
photon events with better energy reconstruction.  We hope
that the evidence presented in this paper for a signal in
general, and specifically the superior value of the
high-$\theta$ events can be used to justify such a change.

Data acquired from the new survey strategy, along with Pass 8 improvements,
will be able to better constrain the morphology of the gamma-ray cusp and
study in detail the line emission in $110 \lesssim
E \lesssim 140$ GeV, determine if there are two (or more) lines, and measure
the ratios of the line strengths.  Such measurements may be the key to
differentiating between WIMP models and deepening our understanding of the
dark sector. 

\emph{Is this a "discovery?": } The local (without trials factor)
significance of our results is high: $5.0\sigma$ for the centered Gaussian 
template fit to energy bins, $5.25\sigma$ for the unbinned Gaussian fit
at $(\ell=-1.5, b=0)$, and $6.5\sigma$ by fitting the NFW
dark matter profile centered at (-1.5, 0) with power index
$\alpha=1.2$, with similar results for an Einasto profile.  Given such high
significances, it would be
tempting to call this detection a "Discovery."  However, we
have a number of concerns:

\begin{itemize}
\item The cusp is off-center by $1.5\degree$ (200 pc) and
this was not expected.  If the theoretical prior against
this is strong, then this reduces confidence in the result
-- or at least a WIMP-related interpretation of the result.
It is now an urgent question for simulators and theorists to
determine whether this offset is unlikely or not in a spiral
Galaxy with a significant bar.
\item The trials factor for this discovery is significant:
The energy part is roughly 300 for a generic line search,
though much less for the $\gamma\gamma$, $\gamma$Z scenario.
The trials factor for the position offset could also be
substantial.
\item Now that we have characterized what looks like a
robust signal, the path to a truly convincing discovery is
clear: modify the survey strategy to accumulate signal on
the GC as quickly as possible, and repeat the experiment
with no trials factor.
\end{itemize}

\vskip 0.15in {\bf \noindent Acknowledgments:} We thank Neal
Weiner, Tracy Slatyer, Christopher Stubbs, Lars
Hernquist, and Michael Kuhlen for helpful discussions. We acknowledge the use
of public data from the \Fermi\ data archive at
\texttt{http://fermi.gsfc.nasa.gov/ssc/}.  This work would
not be possible without the work of hundreds of people, over
many years, to design, build, and operate \Fermi.  M.S. and
D.P.F. are partially supported by the NASA Fermi Guest
Investigator Program.  This research made use of the NASA
Astrophysics Data System (ADS) and the IDL Astronomy User's
Library at Goddard (Available at
\texttt{http://idlastro.gsfc.nasa.gov}).
\bibliography{line}
\bibliographystyle{hapj}

\end{document}